\documentclass[aps, prl, longbibliography, twocolumn,superscriptaddress,preprintnumbers,amsmath,amssymb]{revtex4-2}
\usepackage{appendix}
\usepackage{graphicx}
\usepackage{dcolumn}
\usepackage{bm}
\usepackage{subfigure}
\usepackage{multirow}
\usepackage{amsmath}
\usepackage{color}
\usepackage[
  colorlinks=true,
  urlcolor=blue,
  linkcolor=blue,
  citecolor=blue
]{hyperref}
\definecolor{blue}{rgb}{0,0,1}
\definecolor{red}{rgb}{1,0,0}
\definecolor{green}{rgb}{0,1,0}

\begin{document}

\title{Orientational phase transitions induced by two-patch interactions}

\author{Lingyao Kong}
\affiliation{School of Physics and Optoelectronic Engineering, Anhui University, Hefei 230601, China}

\author{Hua Tong}
\affiliation{Department of Physics, University of Science and Technology of China, Hefei 230026, China}

\author{Hao Hu}
\email[Contact author: ]{huhao@ahu.edu.cn}
\affiliation{School of Physics and Optoelectronic Engineering, Anhui University, Hefei 230601, China}

\begin{abstract}
	For two-patch particles in two dimensions, we find that the coupling of anisotropic patchy interactions and the triangular lattice leads to novel phase behaviors. For asymmetric patch-patch (PP) and nonpatch-nonpatch (NN) interactions, the system has dual orientationally ordered phases of the same symmetry, intermediated by a nematic phase. Both phase transitions from the nematic phase to dual ordered phases are continuous and belong to the same universality class, and they lead to highly nonmonotonic variations of the nematic order parameter.	The system becomes disordered at high temperature through another continuous transition. When the PP and NN interactions become symmetric, the system has subextensive ground-state entropy, and with increasing temperature it undergoes two Berezinskii-Kosterlitz-Thouless phase transitions, with a critical phase connecting a nematic phase and a disordered phase. These results open up new opportunities for designing patchy interactions to study orientational phase transitions and critical phenomena.
\end{abstract}
\maketitle

{\bf Introduction ---}
Due to anisotropic interactions, patchy particles~\cite{Casagrande1989,deGennes}
can assemble into versatile structures~\cite{Zhang2004,Walter2013,Zhang2017,LiDuguet2020,LiJiang2025,He2020,Liu2024,Gregor2024,Beneduce2025}.
For close packed one-patch (also called Janus) particles, the anisotropic interactions give rise to 
transitions between disordered and orientational ordered phases~\cite{Shin2014,Jiang2014,
Mitsumoto2018,Huang2019,Patrykiejew2021,Liang2021,Hu2022,Huang2022},
and critical phenomena similar to those of the classical $3$-state Potts model 
have been reported in two dimensions~\cite{Mitsumoto2018,Patrykiejew2021,Hu2022}.
When Janus particles move in 2D continuum space, a simulation study
suggests that a continuous phase transition still presents at high particle densities~\cite{Liang2021}. 

For two-patch particles, previous studies focused mainly on the assembly of a kagome lattice
in continuum space~\cite{ChenGranick2011,RomanoSciortino2011,Eslami2018,Mallory2019,Bahri2022,Schubert2025}.
The kagome lattice is stabilized by the entropy of both translational and rotational vibrations~\cite{Mao2013} 
and transforms into the triangular lattice at high pressures or densities~\cite{RomanoSciortino2011}.
However, for close-packed two-patch particles, it remains unclear how the coupling of anisotropic interactions
with a lattice affects phase behaviors.
Rich phase transitions and critical phenomena have been found in lattice models of anisotropic interactions,
e.g., compass models describing materials with orbital degrees of freedom~\cite{Nussinov2015},
hard-core spin models~\cite{Placke2023}, pivoted hard disks~\cite{Saryal2023},
nonreciprocal~\cite{Dadhichi2020,Dopierala2025,Popli2025,Loos2023,Bandini2024,Liu2025}
and reciprocal~\cite{Bandini2024} XY models.
Thus, it is tempting to ask if one could observe notable phenomena simply by adding one patch to Janus particles.

In this paper, we demonstrate that the coupling of anisotropic patchy interactions with a lattice
provides a fertile platform for studying orientational phase transitions and critical phenomena.
We couple two-patch interactions with the triangular lattice and observe orientational phase transitions by Monte Carlo (MC) simulations and finite-size scaling (FSS) analysis. 
Each site of the triangular lattice is occupied by a two-patch disk with a diameter of one lattice spacing.
The disk has its center fixed at the lattice site and can only rotate due to coupling
with a thermal reservoir of temperature $T$. Only nearest-neighboring particles interact 
and the interactions can be classified as of patch-patch (PP), nonpatch-nonpatch (NN) and patch-nonpatch (PN) types.
In a model of asymmetric PP and NN interactions --- the interaction strengths being different 
and the patch area being not equal to nonpatch area, we find three continuous phase transitions
when increasing $T$. The first two transitions occur between a six-fold symmetry-broken nematic phase 
and two three-fold symmetry-broken phases, and they are in the same universality class.
Though the dual transitions break a two-fold symmetry, they have critical phenomena different from those of the 2D Ising model.
It is also found that the dual transitions lead to 
highly nonmonotonic variations of the nematic order parameter.
When PP and NN interactions of the two-patch model are symmetric, we find that the ground state
has subextensive entropy associated with an intermediate symmetry ---
being order in one direction and disorder in the other direction~\cite{Nussinov2015},
and breaks a six-fold global symmetry. 
Similar to the $6$-state clock model~\cite{Jose1977,Elitzur1979,Tomita2002,Ueda2020,Chen2022,Tuan2022},
the symmetric two-patch model undergoes two Berezinskii-Kosterlitz-Thouless (BKT) phase transitions, 
i.e., from a low-$T$ nematic phase to an intermediate critical phase,
then to a disordered phase at high $T$. 
More rich critical behaviors are expected when adding more patches
to particles~\cite{Bianchi2006,Tavares2009,Swinkels2024,WangHu2022,WangHu2024}, 
and our results may also help developing other models,
such as hard-core spin models~\cite{Placke2023} and XY models with 
vision core interactions~\cite{Loos2023,Bandini2024,Liu2025}. 

\begin{figure}[htbp] 
   \centering
	\includegraphics[width=3.2 in]{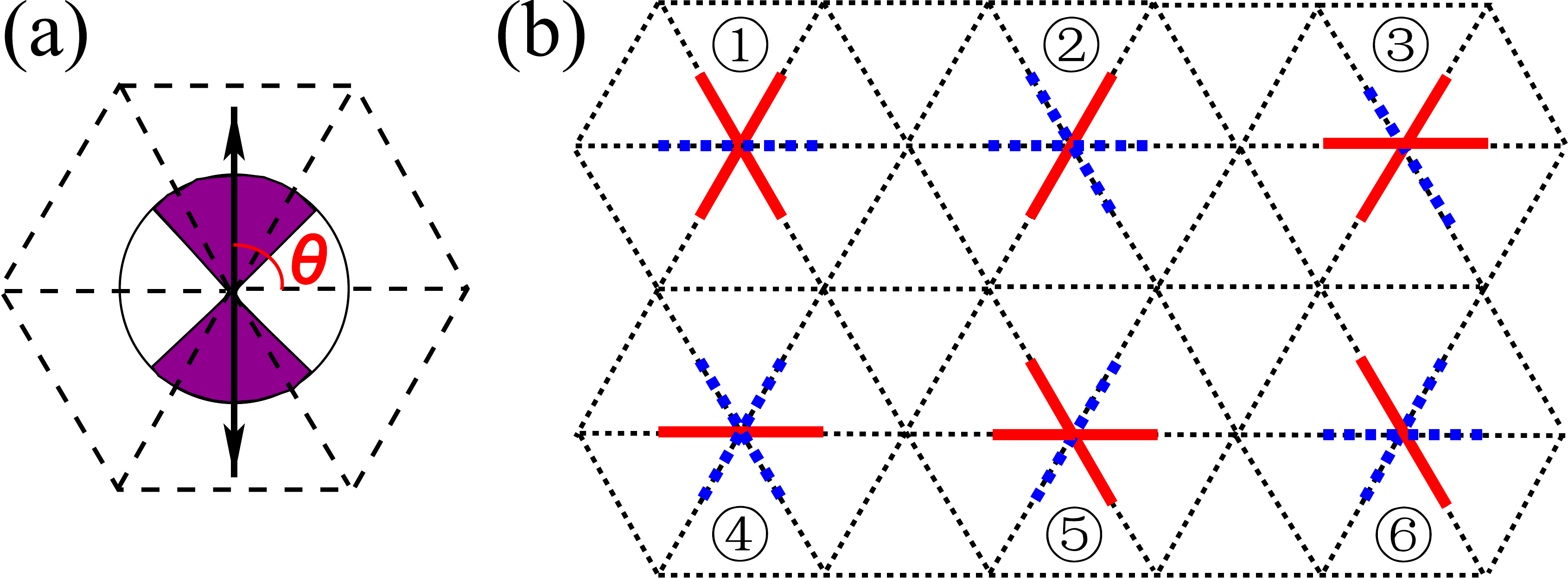}
	\caption{The two-patch model. 
	(a) A disk has two symmetric patches indicated by colored sectors, 
 	 and its orientation is given by a director that 
	 has an angle $\theta$ with respect to the horizontal axis 
	 (or equivalently $\theta+n\pi$, with $n$ being an arbitrary integer). 
	Sizes of patches are characterized the parameter $\chi$, 
	defined as the ratio of the disk covered by patches.
	(b) Six edge-covering states of a two-patch disk with $1/3 < \chi < 1/2$ on the triangular lattice.
    Edge segments covered by patches are labeled as red solid lines, 
	and those covered by nonpatches as  blue dashed lines. 
	The average $\theta$ of directors in states $1$ to $6$ are $\pi/2$, $\pi/3$, $\pi/6$, $0$, $-\pi/6$, $-\pi/3$, respectively.} 
   \label{fig:def_model}
\end{figure}

{\bf Definition of two-patch models ---}
We define the two-patch models on the triangular lattice,
on which each site is occupied by a two-patch disk.
As illustrated in Fig.~\ref{fig:def_model}(a), 
a disk has two equal patches in opposite directions
and its orientation can be characterized by a director. 
A disk interacts with its nearest neighbors through PP, 
NN or PN contacts (NP being equivalent to PN).
The total energy of the system is $E = \sum_{ij} E_{ij}$, 
where $i$ and $j$ denotes nearest-neighboring sites, 
between which the interaction energy $E_{ij}$ depends on the contact type.
Different models are distinguished from each other
by the patchy coverage $\chi$ (the ratio of the disk covered by patches) 
or interaction strengths of contacts.
We assume that interaction strengths only depend on contact types.
In this case, one can reduce the continuous rotational states 
of a disk into several edge-covering states and interactions between disks 
can solely be determined by these states.
For example, when $1/3 < \chi < 2/3$, as shown in Fig~\ref{fig:def_model}(b), 
there exist six edge-covering states, of which three states ($1,3,5$) have four neighboring edges 
being covered by patches, and other three states ($2,4,6$) have two edges covered by patches.
Making a full rotation of the disk, the appearance probability of each state in ($1,3,5$)
is $\chi-1/3$, and of each state in ($2,4,6$) is $2/3 - \chi$. 
In the following we represent the disk by these edge-covering states 
and assume that state $i$ is equivalent to state $i \pm 6$.
We take the orientation of each state to be the average direction of the director 
in the state. 

\begin{figure}[tbph]
   \centering
	\includegraphics[width=1.68 in]{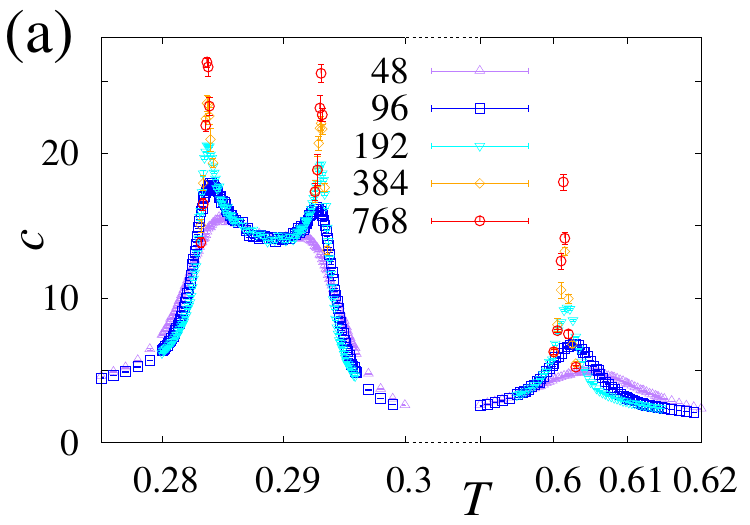}
	\includegraphics[width=1.68 in]{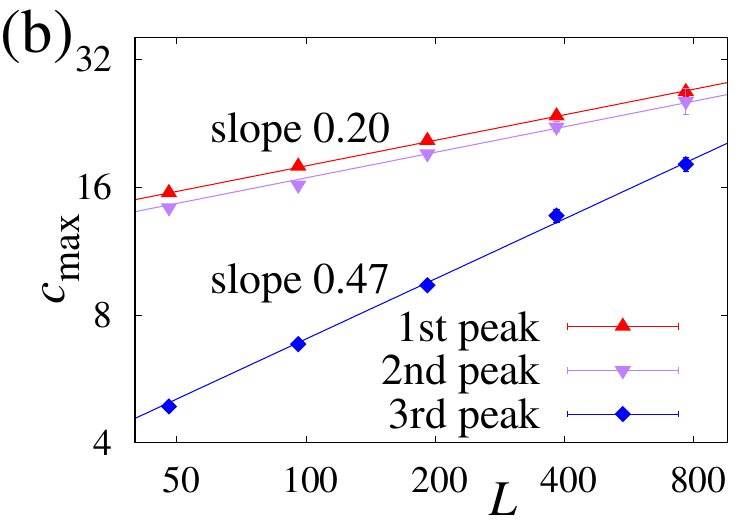}
	\includegraphics[width=3.36 in]{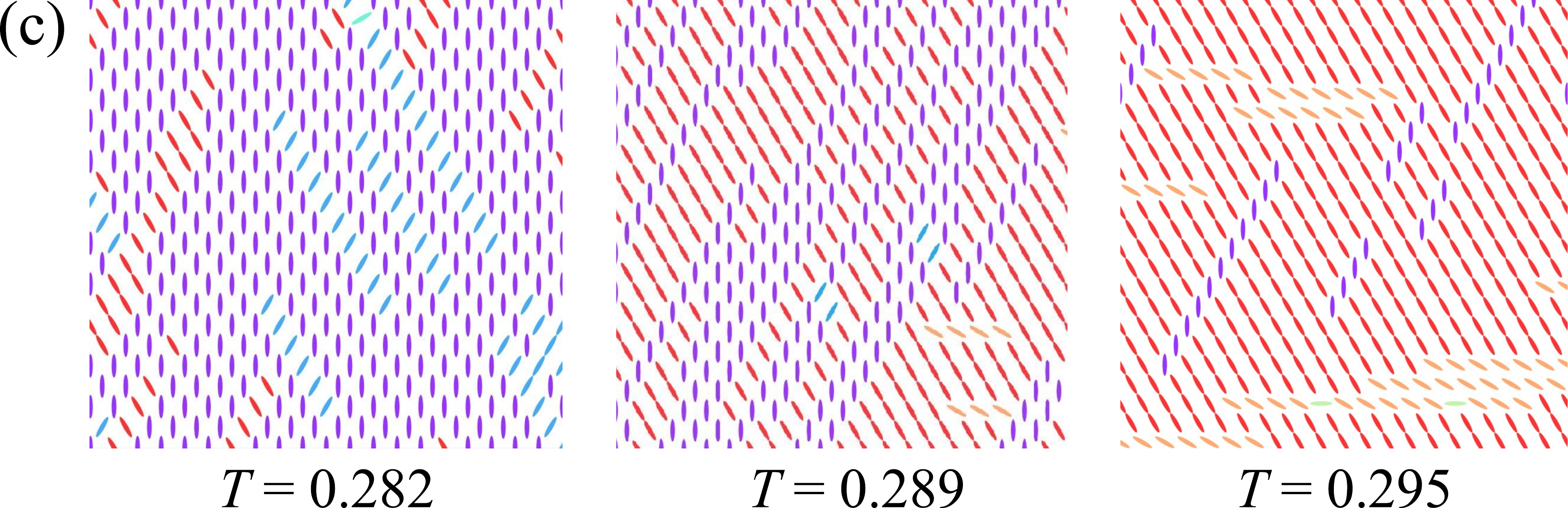}
	\caption{Specific heat and configuration snapshots for the asymmetric model.
	(a) The specific heat $c$ vs. $T$. 
	(b) The maximum of specific heat $c_{\rm max}$ versus $L$ in log-log scales. 
	The three lines correspond to the three peaks from left to right in plot (a).
	(c) Snapshots of the first three phases. Distinct colors label directors 
	in different states. Directors in state $1$ and $6$ dominate
	the first and third snapshots, respectively; and a mixture of 
	the two states dominates the second snapshot.}
   \label{fig:c-asym}
\end{figure}

{\bf Dual phase transitions of the same universality ---}
Phase behaviors of the models are determined by the patchy coverage $\chi$
and interaction strengths of different contact types.
We first consider an asymmetric model
with $\chi=0.4$, $E_{PP}=-1$ and $E_{NN}=-0.6$ and $E_{PN}=0$.
For this model in the ground state, all directors are in the same state, taking
one value out of states ($1,3,5$).
When increasing $T$, we find dual continuous phases transitions
in the same universality class between three ordered phases, 
followed by another continuous phase transition to the fully disordered phase.

To investigate the finite temperature phase behaviors, we performed 
Markov Chain Monte Carlo (MCMC) simulations using the Metropolis algorithm.
The simulations were performed on $L \times L$ triangular lattices
with rhombus-shape periodic boundary conditions.
For $L \le 192$, the simulations were conducted on CPU workstations.
For each simulation job, in one step, a director is randomly selected 
and proposed to rotate to one of other five states besides its current state.
One Monte Carlo sweeps (MCS) contains $L^2$ steps.
For $L \ge 384$, the simulations were conducted on a GPU workstation 
with two NVIDIA GeForce RTX 4090 graphics cards.
For each job, sublattice updates are realized by CUDA programming. 
The triangular lattice can be decomposed into three sublattices. 
In one step, directors on one sublattice are independently proposed to rotate 
to other five states. 
One MCS consists of three sequential sublattice steps.
For each job, the initial configuration were taken as an ordered state
or a last configuration from finished jobs. 
Before sampling, $O(10^7)$ or more MCS were performed for thermalization.
Multiple jobs were performed to improve the statistics.
For sampling near critical points, the total number of MCS was
over $2\times10^8$ for $48 \le L \le 192$, $4\times10^8$ for $L=384$,
and $O(10^9)$ for $L=768$.

\begin{figure*}[!bhtp]
        \includegraphics[height=1.64 in]{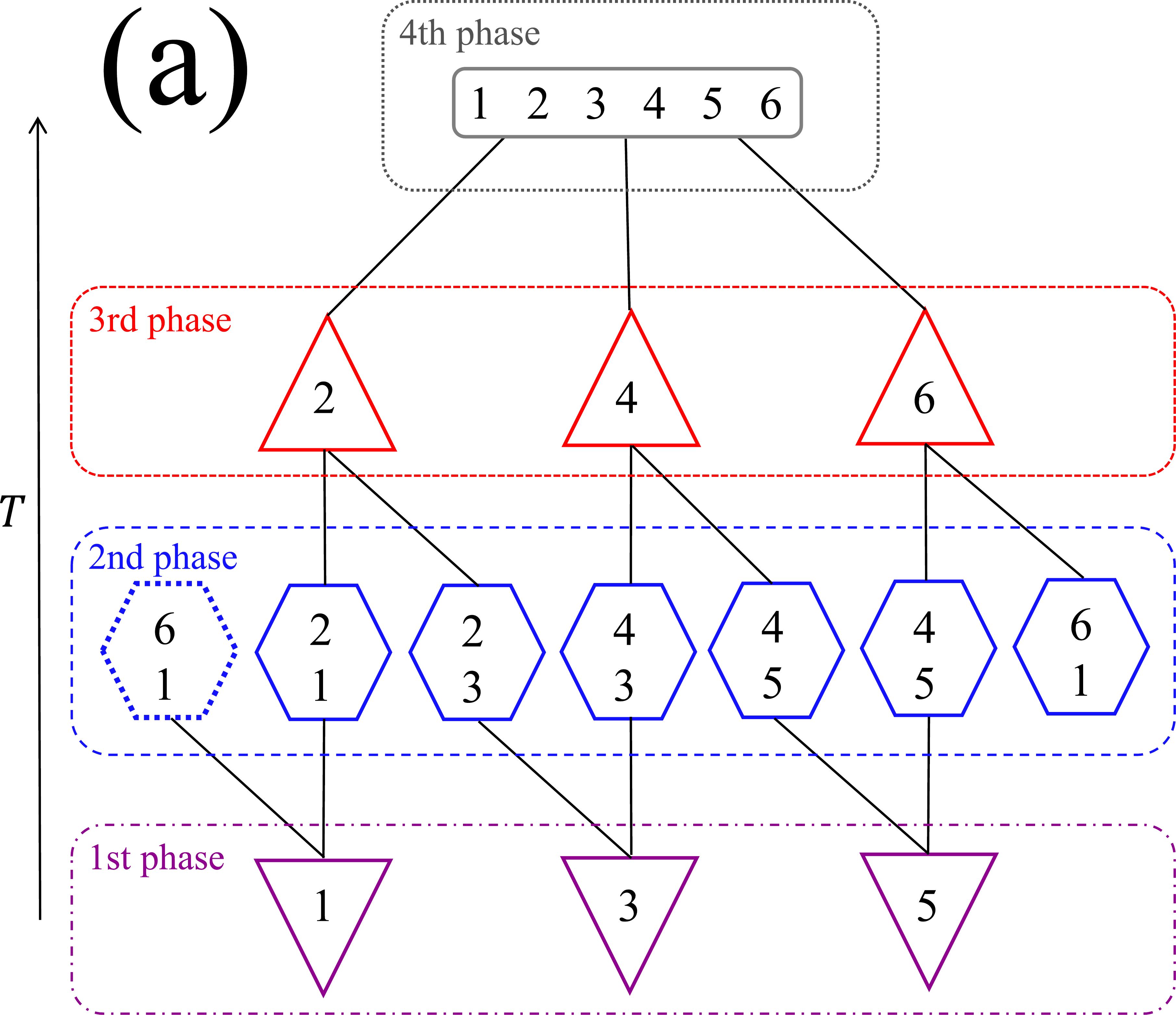}
        \includegraphics[height=1.74 in]{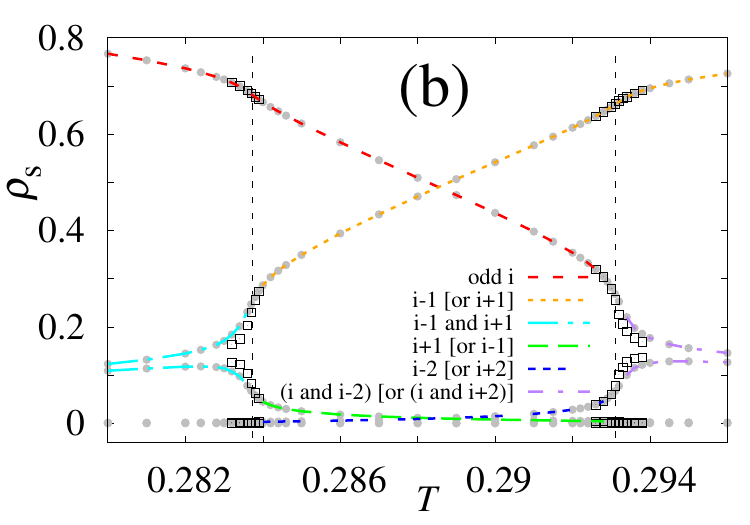}
        \includegraphics[height=1.74 in]{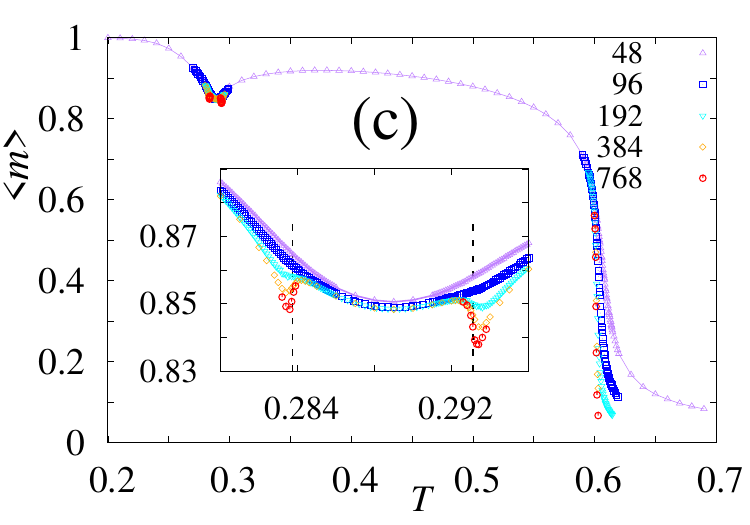}
	\caption{Results for the asymmetric model. 
	(a) Densest director state(s) in four phases. 
        In each phase, states in one polygon dominate a configuration.
	(b) Sorted densities of six states of directors versus $T$.
        Director states of different curves are given by the labels.
	Gray dots represent simulation results at $L=384$,
	and black squares are those at $L=768$.
	An enlargement of the region $\rho_{\rm s}<0.02$ is shown 
	in Fig.~\ref{fig:asym-rhos-enlarge}. 
	(c) The nematic order parameter $<m>$ versus $T$ at different sizes $L$.
	Vertical dotted lines indicate positions of the first two transitions.}
   \label{fig:asym-OP}
\end{figure*}

From simulations results, we plot the specific heat as shown 
in Fig.~\ref{fig:c-asym}(a), where we observe three diverging peaks
which suggest three phase transitions. 
FSS theory predicts that for continuous phase transitions
the peak height scales as $c_{\rm max}(L) \sim L^{-d+2/\nu}=L^{-2+2/\nu}$, 
and the peak position scales as $T_c(L) - T_c(\infty) \sim L^{-\lambda}$~\cite{Barber1983}.
The peak heights of the three transitions are plotted in Fig.~\ref{fig:c-asym}(b), 
from which one gets $1/\nu = 1.10(1)$ for the first two transitions,
and $1/\nu = 1.24(2)$ for the third transition.
Moreover, from the peak positions $T_c(L)$ in Fig.~\ref{fig:asym-TcL},
we estimate the critical temperatures as $T_{c1} = 0.28374(5)$, $T_{c2} \simeq 0.29310(1)$
and $T_{c3} \simeq 0.60114(9)$, and the shift exponent $\lambda$
as $2.1(4)$, $2.3(1)$ and $1.1(1)$, respectively. 
Markedly the first two transitions have $\lambda \ne 1/\nu$, 
which reminds that ``$\lambda = 1/\nu$ is not a necessary conclusion 
of finite-size scaling~\cite{Barber1983}."
These results demonstrate that the three phase transitions are continuous, 
and that the first two transitions are in the same universality class.

To explore in depth phase transitions of the asymmetric model, 
we plot snapshots of the first three phases in Fig.~\ref{fig:c-asym}(c) 
and summarize the densest director state(s) in Fig.~\ref{fig:asym-OP}(a).
In the first phase, one director state out of states $(1,3,5)$
dominates a configuration;
and in the third phase one state out of $(2,4,6)$ dominates.
Both the first and the third phases break a three-fold symmetry.
The second phase is an intermediate phase connecting the above two symmetry-breaking phases,
and it resembles a nematic phase --- the directors assemble into 
aligned chains with random positions.  
In this phase, an odd state $i$ first dominates a configuration,
then the dominating state crossovers to an even state $j=i+1$ or $i-1$ as $T$ increases. 
There are six possible pairs of $(i,j)$, thus the second phase breaks six-fold symmetry.
In the fourth phase, the system is disordered and the six states
of directors appear with same probabilities in a configuration.
From the first and third phases to the second phase, the change of symmetry is the same,
which explains our numerical observation of same universality for the first two phase transitions. 

The above scenario is quantitatively supported by numerical results for 
sorted densities $\rho_{\rm s}$ of director states in Fig.~\ref{fig:asym-OP}(b).
To obtain the order parameter $\rho_{\rm s}(1:6)$, we first measured for each configuration 
the densities of six director states, then sorted the six densities, 
and finally calculated averages over many configurations. 
In the first phase, the maximum density $\rho_{\rm s}(1)$ corresponds to directors of an odd state $i$,
the second and third largest densities $\rho_{\rm s}(2:3)$ correspond to even director states $i-1$ and $i+1$
and the two densities approach each other as the system size increases.
At the first phase transition, a two-fold symmetry between $i-1$ and $i+1$ is broken,
while no obvious change occurs for state $i$.
In other words, the state $i$ directors correlate directors of states $i-1$ and $i+1$, 
and they serve as an anisotropic background for the two-fold symmetry breaking. 
These may explain the inequality $\lambda \ne 1/\nu$~\cite{Chen2019} 
and why critical behaviors are different from those of the 2D Ising model ($1/\nu=1$),
which require further investigation.
The densities $\rho_{\rm s}(4:5)$ behave similarly to $\rho_{\rm s}(2:3)$,
as shown in Fig.~\ref{fig:asym-rhos-enlarge}. 
Similar behaviors are observed at the second phase transition.
For the second phase, crossover behaviors are found near $T=0.2885$ for $\rho_{\rm s}$
in Figs.~\ref{fig:asym-OP}(b) and \ref{fig:asym-rhos-enlarge}, which are crucial for 
the second phase's role of connecting the first and third phases.
For the third phase transition, since there is a three-fold symmetry breaking from the fourth to the third phase, 
similar to phase transitions for Janus particles~\cite{Mitsumoto2018,Hu2022} 
or interacting rigid rods~\cite{dosSantos2023} on the triangular lattice,
values of critical exponents are close to those of the $3$-state Potts model ($1/\nu=6/5$).

\begin{figure*}[!bhtp]
        \resizebox{180mm}{!}{\includegraphics[trim=0.0in 0.0in 0.0in 0.0in]{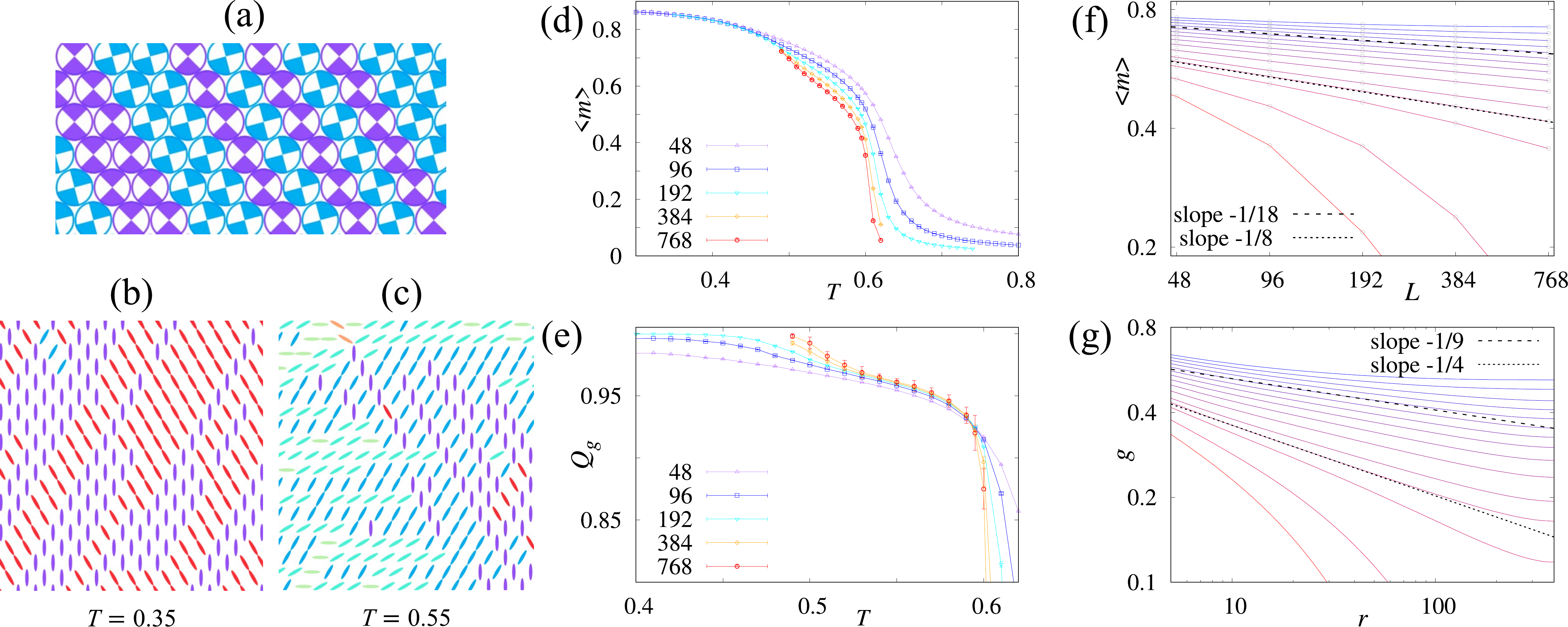}}
	\caption{Results for the symmetric model.
	(a) Snapshot of a ground state, which consisting of disks in states $1$ and $2$. 
	Patches for the two states are colored purple and blue respectively.
	(b) Snapshot of a configuration at ($T=0.35$, $L=192$) in the low-$T$ ordered phase.
	(c) Snapshot of a configuration at ($T=0.55$, $L=192$) in the intermediate critical phase.
	(d) The nematic order parameter $\langle m \rangle$ versus $T$.
	(e) The correlation ratio $Q_g$ versus $T$.
	(f) The nematic order parameter $\langle m \rangle$ versus $L$.
	(g) The nematic correlation function $g(r)$ at $L=768$.
	For (f) and (g), from top to bottom, the temperature increases from $T=0.49$ to $0.62$  
	with $\Delta T=0.01$, except for the curve near the dotted line which has $T=0.595$. 
	}
   \label{fig:sym}
\end{figure*}

To further characterize the order of the phases, we define
\begin{equation}
	{m}^2 = | \frac{1}{N} \sum_{i=1}^{N} \cos(2\theta_i) |^2 + | \frac{1}{N} \sum_{i=1}^{N} \sin(2\theta_i) |^2 \;.
\end{equation}
and measure the nematic order parameter $\langle m \rangle$.
The results are plotted in Fig.~\ref{fig:asym-OP}(c).
While the fourth phase is disordered with $\langle m \rangle$ approaches zero for large $L$, 
it appears interesting to see highly nonmonotonic continuous variation of 
nonzero $\langle m \rangle$ versus $T$ in the first three phases. 
The latter can be understood as follows.
Increasing $T$, in the first phase, the decreasing of $\langle m \rangle$ is 
related to the weakening dominance of an odd state $i$ 
and the strengthening appearance of two neighboring states $i-1$ and $i+1$.
In the second phase, the initial increasing of $\langle m \rangle$
is related to the gradually taken over of even state $i-1$ (or $i+1$) 
against the other state $i+1$ (or $i-1$); the following decreasing 
of $\langle m \rangle$ is related to that the odd state $i$ gradually lose its dominance
to the even state $i-1$ (or $i+1$) till the minimum of $\langle m \rangle$. 
Similar analyses explain the following change of $\langle m \rangle$ from the second to the third phase.
Future work may explore how the nonmonotonic behavior affects applications, e.g., transport properties.

{\bf Double BKT transitions ---}
If varying the coverage $\chi$ or interaction strengths,
the stability of phases would change.
In the following we show that in a symmetric model 
there is a quasi-long-range ordered (critical) phase, connecting a low-$T$ nematic
phase and a high-$T$ disordered phase through two transitions of BKT-type.

The symmetric model has $\chi=1/2$, $E_{PP}=E_{NN}=-1$ and $E_{PN}=0$.
Figure~\ref{fig:sym}(a) exemplifies that two neighboring director states 
make a kind of ground states, which has a degeneracy of $2^L$.  
Six different pairs of neighboring director states make six kinds of ground states. 
The entropy of ground states is $S = k \ln{(6 \times 2^L)} = k L \ln{2} + k \ln{6}$,
where $k$ is the Boltzmann constant.
Thus $S/L^d \sim L^{1-d} = L^{-1}$, i.e., the ground-state entropy is subextensive.
The ground state has an intermediate symmetry~\cite{Nussinov2015}, i.e., it is ordered in the direction 
along the stripes but disordered in the direction perpendicular to the stripes. 
Here ``disordered" means that a stripe of width one can assume
either of the two director states.  Increasing $T$ from zero, in the thermodynamic limit $L \to \infty$, 
the subsystems along the stripes become disordered, since a stripe of width one 
is similar to the one-dimensional (1D) Ising model, for which a single kink can destroy the 1D order,
as shown in Fig.~\ref{fig:sym}(b).
However, at low temperatures, the system can still be regarded as ``ordered",
in the sense that one pair of neighboring director states out of six possible pairs dominates 
a configuration, i.e., a six-fold global symmetry is broken.

For simulations of the symmetric model, $O(10^7)$ or more MCS were performed for thermalization.
In the sampling stage for each temperature $0.49 \le T \le 0.62$, in total over $2 \times 10^8$ MCS 
were performed for each $L \le 192$, and the number was over $10^9$ MCS for $L = 384$ and $768$.
Results of $\langle m \rangle$ in a wide range of $T$ are plotted in Fig.~\ref{fig:sym}(d).
It is seen that $\langle m \rangle >0$ when $T<T_{c1} \simeq 0.50$,
$\langle m \rangle$ slowly decreases for increasing $L$ when $T_{c1} < T < T_{c2} \simeq 0.60$, 
and $\langle m \rangle$ quickly tends to zero for increasing $L$ when $T>T_{c2}$.
These behaviors of the order parameter resemble those of 
the classical $q$-state clock models with $q \ge 5$~\cite{Tomita2002,Borisenko2011,Ueda2020},
suggesting double BKT phase transitions and  a critical phase for $T_{c1} < T < T_{c2}$,
though the ground states of the current model is very different from those of the clock model. 
A snapshot of the intermediate critical phase is shown in Fig.~\ref{fig:sym}(c),
where no explicit long-range order is observed.

To further investigate the BKT behaviors,
we define the nematic correlation 
\begin{equation}
	g(r) = \langle \cos{2(\theta_i-\theta_j)} \delta(|\vec{r}_i - \vec{r}_j|-r) \rangle
	\label{def:gr}
\end{equation}
The correlations $g(L/2)$ and $g(L/4)$ were measured, as well as a dimensionless ratio
$Q_g(L)= g(L/2)/g(L/4)$~\cite{Surungan2019,Okabe2025}.
For $q$-state clock models, $Q_g$ has been found to be exceptionally suitable for determining double BKT phase transitions~\cite{Surungan2019,Okabe2025}.
Figure~\ref{fig:sym}(e) demonstrates that $Q_g$ also works quite well for the symmetric model. 
From the plot, more accurate estimates of the transition temperatures are
estimated as $T_{c1} = 0.530(4)$ and $T_{c2} = 0.595(4)$, 
where curves of large sizes start deviating from each other.
The collapse of large-size $Q_g$ curves for $T_{c1} < T < T_{c2}$ is another evidence
that the intermediate phase is critical.
We also measured $g(r)$ for different temperatures
at $L=768$, as plotted in Fig.~\ref{fig:sym}(g).
It is found that $g(r)$ is long-range for $T < T_{c1}$,
quasi-long-range for $T_{c1} < T < T_{c2}$, short-range for $T > T_{c2}$.
These confirm that the intermediate phase is critical.
At criticality one has $g(r) \sim r^{-\eta}$, where $\eta$ is the anomalous exponent.
Results in Fig.~\ref{fig:sym}(g) show that $\eta$ changes from $1/9$ near $T_{c1}$
to $1/4$ near $T_{c2}$ when increasing $T$.
FSS predicts that at criticality $ \langle m \rangle \sim L^{(2-d-\eta)/2} = L^{-\eta/2}$.
Our results of $\langle m \rangle$ versus $L$ in Fig.~\ref{fig:sym}(f) also show 
that $\eta$ increases from $1/9$ near $T_{c1}$ to $1/4$ near $T_{c2}$ when increasing $T$.
These values of $\eta$ strongly suggest that the symmetric model 
and the $6$-state clock model share the same universality class~\cite{Jose1977,Elitzur1979}.

{\bf Summary and outlook} ---
We find that, after adding one patch to Janus particles, the coupling of anisotropic two-patch interactions 
with a triangular lattice leads to rich phase behaviors, including dual continuous phase transitions in the same universality class
between three orientationally ordered phases for asymmetric PP and NN interactions, 
and double BKT phase transitions for symmetric interactions. 
Varying interactions, future work may explore the crossover from the asymmetric
to the symmetric model.
Specific interactions in two-patch models could lead to geometric frustration, 
e.g., on the triangular lattice frustration occurs in an asymmetric model
with $\chi<1/3$, $E_{PP}=1$, $E_{PN}=-1$ and $E_{NN}=0$~\cite{Ding2025}.
For Janus particles in 2D continuum space, simulation results suggest that 
a continuous phase transition presents at high densities~\cite{Liang2021},
while at lower densities the coupling between translational and orientational motions 
dramatically changes the phase behaviors~\cite{Liang2021,Huang2022}.
It would be interesting to explore how the observed phases behaviors change 
when two-patch particles move in 2D continuum space.
Study of two-patch models in nonequilibrium conditions also deserves more attention~\cite{Mallory2019,Landi2025,Schubert2025}.
Experimentally, one could expect our findings be observed in colloidal systems~\cite{ChenGranick2011}
or rotating magnetic units~\cite{Du2022,Wang2025}. 

\emph{Acknowledgments}: 
We thank Qunli Lei for helpful discussions. 
This work has been supported by the National Natural Science Foundation of China under grant No.~12375026.
We acknowledge the High-Performance Computing Platform of Anhui University for providing computing resources.

\bibliography{patchyPT.bib}

\begin{thebibliography}{56}%
\makeatletter
\providecommand \@ifxundefined [1]{%
 \@ifx{#1\undefined}
}%
\providecommand \@ifnum [1]{%
 \ifnum #1\expandafter \@firstoftwo
 \else \expandafter \@secondoftwo
 \fi
}%
\providecommand \@ifx [1]{%
 \ifx #1\expandafter \@firstoftwo
 \else \expandafter \@secondoftwo
 \fi
}%
\providecommand \natexlab [1]{#1}%
\providecommand \enquote  [1]{``#1''}%
\providecommand \bibnamefont  [1]{#1}%
\providecommand \bibfnamefont [1]{#1}%
\providecommand \citenamefont [1]{#1}%
\providecommand \href@noop [0]{\@secondoftwo}%
\providecommand \href [0]{\begingroup \@sanitize@url \@href}%
\providecommand \@href[1]{\@@startlink{#1}\@@href}%
\providecommand \@@href[1]{\endgroup#1\@@endlink}%
\providecommand \@sanitize@url [0]{\catcode `\\12\catcode `\$12\catcode `\&12\catcode `\#12\catcode `\^12\catcode `\_12\catcode `\%12\relax}%
\providecommand \@@startlink[1]{}%
\providecommand \@@endlink[0]{}%
\providecommand \url  [0]{\begingroup\@sanitize@url \@url }%
\providecommand \@url [1]{\endgroup\@href {#1}{\urlprefix }}%
\providecommand \urlprefix  [0]{URL }%
\providecommand \Eprint [0]{\href }%
\providecommand \doibase [0]{https://doi.org/}%
\providecommand \selectlanguage [0]{\@gobble}%
\providecommand \bibinfo  [0]{\@secondoftwo}%
\providecommand \bibfield  [0]{\@secondoftwo}%
\providecommand \translation [1]{[#1]}%
\providecommand \BibitemOpen [0]{}%
\providecommand \bibitemStop [0]{}%
\providecommand \bibitemNoStop [0]{.\EOS\space}%
\providecommand \EOS [0]{\spacefactor3000\relax}%
\providecommand \BibitemShut  [1]{\csname bibitem#1\endcsname}%
\let\auto@bib@innerbib\@empty
\bibitem [{\citenamefont {Casagrande}\ \emph {et~al.}(1989)\citenamefont {Casagrande}, \citenamefont {Fabre}, \citenamefont {Rapha\"el},\ and\ \citenamefont {Veyssi\'e}}]{Casagrande1989}%
  \BibitemOpen
  \bibfield  {author} {\bibinfo {author} {\bibfnamefont {C.}~\bibnamefont {Casagrande}}, \bibinfo {author} {\bibfnamefont {P.}~\bibnamefont {Fabre}}, \bibinfo {author} {\bibfnamefont {E.}~\bibnamefont {Rapha\"el}},\ and\ \bibinfo {author} {\bibfnamefont {M.}~\bibnamefont {Veyssi\'e}},\ }\bibfield  {title} {\bibinfo {title} {``{J}anus beads": Realization and behaviour at water/oil interfaces},\ }\href {https://doi.org/10.1209/0295-5075/9/3/011} {\bibfield  {journal} {\bibinfo  {journal} {Europhysics Letters}\ }\textbf {\bibinfo {volume} {9}},\ \bibinfo {pages} {251} (\bibinfo {year} {1989})}\BibitemShut {NoStop}%
\bibitem [{\citenamefont {de~Gennes}(1992)}]{deGennes}%
  \BibitemOpen
  \bibfield  {author} {\bibinfo {author} {\bibfnamefont {P.~G.}\ \bibnamefont {de~Gennes}},\ }\bibfield  {title} {\bibinfo {title} {Soft matter},\ }\href {https://doi.org/10.1103/RevModPhys.64.645} {\bibfield  {journal} {\bibinfo  {journal} {Rev. Mod. Phys.}\ }\textbf {\bibinfo {volume} {64}},\ \bibinfo {pages} {645} (\bibinfo {year} {1992})}\BibitemShut {NoStop}%
\bibitem [{\citenamefont {Zhang}\ and\ \citenamefont {Glotzer}(2004)}]{Zhang2004}%
  \BibitemOpen
  \bibfield  {author} {\bibinfo {author} {\bibfnamefont {Z.}~\bibnamefont {Zhang}}\ and\ \bibinfo {author} {\bibfnamefont {S.~C.}\ \bibnamefont {Glotzer}},\ }\bibfield  {title} {\bibinfo {title} {Self-assembly of patchy particles},\ }\href {https://doi.org/10.1021/nl0493500} {\bibfield  {journal} {\bibinfo  {journal} {Nano Letters}\ }\textbf {\bibinfo {volume} {4}},\ \bibinfo {pages} {1407} (\bibinfo {year} {2004})}\BibitemShut {NoStop}%
\bibitem [{\citenamefont {Walther}\ and\ \citenamefont {M{\"u}ller}(2013)}]{Walter2013}%
  \BibitemOpen
  \bibfield  {author} {\bibinfo {author} {\bibfnamefont {A.}~\bibnamefont {Walther}}\ and\ \bibinfo {author} {\bibfnamefont {A.~H.~E.}\ \bibnamefont {M{\"u}ller}},\ }\bibfield  {title} {\bibinfo {title} {Janus particles: Synthesis, self-assembly, physical properties, and applications},\ }\href {https://doi.org/10.1021/cr300089t} {\bibfield  {journal} {\bibinfo  {journal} {Chemical Reviews}\ }\textbf {\bibinfo {volume} {113}},\ \bibinfo {pages} {5194} (\bibinfo {year} {2013})}\BibitemShut {NoStop}%
\bibitem [{\citenamefont {Zhang}\ \emph {et~al.}(2017)\citenamefont {Zhang}, \citenamefont {Grzybowski},\ and\ \citenamefont {Granick}}]{Zhang2017}%
  \BibitemOpen
  \bibfield  {author} {\bibinfo {author} {\bibfnamefont {J.}~\bibnamefont {Zhang}}, \bibinfo {author} {\bibfnamefont {B.~A.}\ \bibnamefont {Grzybowski}},\ and\ \bibinfo {author} {\bibfnamefont {S.}~\bibnamefont {Granick}},\ }\bibfield  {title} {\bibinfo {title} {Janus particle synthesis, assembly, and application},\ }\href {https://doi.org/10.1021/acs.langmuir.7b01123} {\bibfield  {journal} {\bibinfo  {journal} {Langmuir}\ }\textbf {\bibinfo {volume} {33}},\ \bibinfo {pages} {6964} (\bibinfo {year} {2017})}\BibitemShut {NoStop}%
\bibitem [{\citenamefont {Li}\ \emph {et~al.}(2020)\citenamefont {Li}, \citenamefont {Palis}, \citenamefont {Mérindol}, \citenamefont {Majimel}, \citenamefont {Ravaine},\ and\ \citenamefont {Duguet}}]{LiDuguet2020}%
  \BibitemOpen
  \bibfield  {author} {\bibinfo {author} {\bibfnamefont {W.}~\bibnamefont {Li}}, \bibinfo {author} {\bibfnamefont {H.}~\bibnamefont {Palis}}, \bibinfo {author} {\bibfnamefont {R.}~\bibnamefont {Mérindol}}, \bibinfo {author} {\bibfnamefont {J.}~\bibnamefont {Majimel}}, \bibinfo {author} {\bibfnamefont {S.}~\bibnamefont {Ravaine}},\ and\ \bibinfo {author} {\bibfnamefont {E.}~\bibnamefont {Duguet}},\ }\bibfield  {title} {\bibinfo {title} {Colloidal molecules and patchy particles: complementary concepts{,} synthesis and self-assembly},\ }\href {https://doi.org/10.1039/C9CS00804G} {\bibfield  {journal} {\bibinfo  {journal} {Chem. Soc. Rev.}\ }\textbf {\bibinfo {volume} {49}},\ \bibinfo {pages} {1955} (\bibinfo {year} {2020})}\BibitemShut {NoStop}%
\bibitem [{\citenamefont {Li}\ \emph {et~al.}(2025)\citenamefont {Li}, \citenamefont {Liu}, \citenamefont {Demirci}, \citenamefont {Dey}, \citenamefont {Rawah}, \citenamefont {Chaudary}, \citenamefont {Ortega}, \citenamefont {Yang}, \citenamefont {Pirhadi}, \citenamefont {Huang}, \citenamefont {Yong},\ and\ \citenamefont {Jiang}}]{LiJiang2025}%
  \BibitemOpen
  \bibfield  {author} {\bibinfo {author} {\bibfnamefont {Y.}~\bibnamefont {Li}}, \bibinfo {author} {\bibfnamefont {F.}~\bibnamefont {Liu}}, \bibinfo {author} {\bibfnamefont {S.}~\bibnamefont {Demirci}}, \bibinfo {author} {\bibfnamefont {U.~K.}\ \bibnamefont {Dey}}, \bibinfo {author} {\bibfnamefont {T.}~\bibnamefont {Rawah}}, \bibinfo {author} {\bibfnamefont {A.}~\bibnamefont {Chaudary}}, \bibinfo {author} {\bibfnamefont {R.}~\bibnamefont {Ortega}}, \bibinfo {author} {\bibfnamefont {Z.}~\bibnamefont {Yang}}, \bibinfo {author} {\bibfnamefont {E.}~\bibnamefont {Pirhadi}}, \bibinfo {author} {\bibfnamefont {B.}~\bibnamefont {Huang}}, \bibinfo {author} {\bibfnamefont {X.}~\bibnamefont {Yong}},\ and\ \bibinfo {author} {\bibfnamefont {S.}~\bibnamefont {Jiang}},\ }\bibfield  {title} {\bibinfo {title} {Two sides of the coin: synthesis and applications of {J}anus particles},\ }\href {https://doi.org/10.1039/D4NR03652B} {\bibfield  {journal} {\bibinfo  {journal} {Nanoscale}\ }\textbf {\bibinfo {volume} {17}},\ \bibinfo
  {pages} {88} (\bibinfo {year} {2025})}\BibitemShut {NoStop}%
\bibitem [{\citenamefont {He}\ \emph {et~al.}(2020)\citenamefont {He}, \citenamefont {Gales}, \citenamefont {Ducrot}, \citenamefont {Gong}, \citenamefont {Yi}, \citenamefont {Sacanna},\ and\ \citenamefont {Pine}}]{He2020}%
  \BibitemOpen
  \bibfield  {author} {\bibinfo {author} {\bibfnamefont {M.}~\bibnamefont {He}}, \bibinfo {author} {\bibfnamefont {J.~P.}\ \bibnamefont {Gales}}, \bibinfo {author} {\bibfnamefont {{\'E}.}~\bibnamefont {Ducrot}}, \bibinfo {author} {\bibfnamefont {Z.}~\bibnamefont {Gong}}, \bibinfo {author} {\bibfnamefont {G.-R.}\ \bibnamefont {Yi}}, \bibinfo {author} {\bibfnamefont {S.}~\bibnamefont {Sacanna}},\ and\ \bibinfo {author} {\bibfnamefont {D.~J.}\ \bibnamefont {Pine}},\ }\bibfield  {title} {\bibinfo {title} {Colloidal diamond},\ }\href {https://doi.org/10.1038/s41586-020-2718-6} {\bibfield  {journal} {\bibinfo  {journal} {Nature}\ }\textbf {\bibinfo {volume} {585}},\ \bibinfo {pages} {524} (\bibinfo {year} {2020})}\BibitemShut {NoStop}%
\bibitem [{\citenamefont {Liu}\ \emph {et~al.}(2024)\citenamefont {Liu}, \citenamefont {Matthies}, \citenamefont {Russo}, \citenamefont {Rovigatti}, \citenamefont {Narayanan}, \citenamefont {Diep}, \citenamefont {McKeen}, \citenamefont {Gang}, \citenamefont {Stephanopoulos}, \citenamefont {Sciortino}, \citenamefont {Yan}, \citenamefont {Romano},\ and\ \citenamefont {{\v{S}}ulc}}]{Liu2024}%
  \BibitemOpen
  \bibfield  {author} {\bibinfo {author} {\bibfnamefont {H.}~\bibnamefont {Liu}}, \bibinfo {author} {\bibfnamefont {M.}~\bibnamefont {Matthies}}, \bibinfo {author} {\bibfnamefont {J.}~\bibnamefont {Russo}}, \bibinfo {author} {\bibfnamefont {L.}~\bibnamefont {Rovigatti}}, \bibinfo {author} {\bibfnamefont {R.~P.}\ \bibnamefont {Narayanan}}, \bibinfo {author} {\bibfnamefont {T.}~\bibnamefont {Diep}}, \bibinfo {author} {\bibfnamefont {D.}~\bibnamefont {McKeen}}, \bibinfo {author} {\bibfnamefont {O.}~\bibnamefont {Gang}}, \bibinfo {author} {\bibfnamefont {N.}~\bibnamefont {Stephanopoulos}}, \bibinfo {author} {\bibfnamefont {F.}~\bibnamefont {Sciortino}}, \bibinfo {author} {\bibfnamefont {H.}~\bibnamefont {Yan}}, \bibinfo {author} {\bibfnamefont {F.}~\bibnamefont {Romano}},\ and\ \bibinfo {author} {\bibfnamefont {P.}~\bibnamefont {{\v{S}}ulc}},\ }\bibfield  {title} {\bibinfo {title} {Inverse design of a pyrochlore lattice of {DNA} origami through model-driven experiments},\ }\href
  {https://doi.org/10.1126/science.adl5549} {\bibfield  {journal} {\bibinfo  {journal} {Science}\ }\textbf {\bibinfo {volume} {384}},\ \bibinfo {pages} {776} (\bibinfo {year} {2024})}\BibitemShut {NoStop}%
\bibitem [{\citenamefont {Posnjak}\ \emph {et~al.}(2024)\citenamefont {Posnjak}, \citenamefont {Yin}, \citenamefont {Butler}, \citenamefont {Bienek}, \citenamefont {Dass}, \citenamefont {Lee}, \citenamefont {Sharp},\ and\ \citenamefont {Liedl}}]{Gregor2024}%
  \BibitemOpen
  \bibfield  {author} {\bibinfo {author} {\bibfnamefont {G.}~\bibnamefont {Posnjak}}, \bibinfo {author} {\bibfnamefont {X.}~\bibnamefont {Yin}}, \bibinfo {author} {\bibfnamefont {P.}~\bibnamefont {Butler}}, \bibinfo {author} {\bibfnamefont {O.}~\bibnamefont {Bienek}}, \bibinfo {author} {\bibfnamefont {M.}~\bibnamefont {Dass}}, \bibinfo {author} {\bibfnamefont {S.}~\bibnamefont {Lee}}, \bibinfo {author} {\bibfnamefont {I.~D.}\ \bibnamefont {Sharp}},\ and\ \bibinfo {author} {\bibfnamefont {T.}~\bibnamefont {Liedl}},\ }\bibfield  {title} {\bibinfo {title} {Diamond-lattice photonic crystals assembled from {DNA} origami},\ }\href {https://doi.org/10.1126/science.adl2733} {\bibfield  {journal} {\bibinfo  {journal} {Science}\ }\textbf {\bibinfo {volume} {384}},\ \bibinfo {pages} {781} (\bibinfo {year} {2024})}\BibitemShut {NoStop}%
\bibitem [{\citenamefont {Beneduce}\ \emph {et~al.}(2025)\citenamefont {Beneduce}, \citenamefont {Pinto}, \citenamefont {Rovigatti}, \citenamefont {Romano}, \citenamefont {\ifmmode~\check{S}\else \v{S}\fi{}ulc}, \citenamefont {Sciortino},\ and\ \citenamefont {Russo}}]{Beneduce2025}%
  \BibitemOpen
  \bibfield  {author} {\bibinfo {author} {\bibfnamefont {C.}~\bibnamefont {Beneduce}}, \bibinfo {author} {\bibfnamefont {D.~E.~P.}\ \bibnamefont {Pinto}}, \bibinfo {author} {\bibfnamefont {L.}~\bibnamefont {Rovigatti}}, \bibinfo {author} {\bibfnamefont {F.}~\bibnamefont {Romano}}, \bibinfo {author} {\bibfnamefont {P.}~\bibnamefont {\ifmmode~\check{S}\else \v{S}\fi{}ulc}}, \bibinfo {author} {\bibfnamefont {F.}~\bibnamefont {Sciortino}},\ and\ \bibinfo {author} {\bibfnamefont {J.}~\bibnamefont {Russo}},\ }\bibfield  {title} {\bibinfo {title} {Falsifiability test for classical nucleation theory},\ }\href {https://doi.org/10.1103/PhysRevLett.134.148201} {\bibfield  {journal} {\bibinfo  {journal} {Phys. Rev. Lett.}\ }\textbf {\bibinfo {volume} {134}},\ \bibinfo {pages} {148201} (\bibinfo {year} {2025})}\BibitemShut {NoStop}%
\bibitem [{\citenamefont {Shin}\ and\ \citenamefont {Schweizer}(2014)}]{Shin2014}%
  \BibitemOpen
  \bibfield  {author} {\bibinfo {author} {\bibfnamefont {H.}~\bibnamefont {Shin}}\ and\ \bibinfo {author} {\bibfnamefont {K.~S.}\ \bibnamefont {Schweizer}},\ }\bibfield  {title} {\bibinfo {title} {Theory of two-dimensional self-assembly of {J}anus colloids: crystallization and orientational ordering},\ }\href {https://doi.org/10.1039/C3SM52094C} {\bibfield  {journal} {\bibinfo  {journal} {Soft Matter}\ }\textbf {\bibinfo {volume} {10}},\ \bibinfo {pages} {262} (\bibinfo {year} {2014})}\BibitemShut {NoStop}%
\bibitem [{\citenamefont {Jiang}\ \emph {et~al.}(2014)\citenamefont {Jiang}, \citenamefont {Yan}, \citenamefont {Whitmer}, \citenamefont {Anthony}, \citenamefont {Luijten},\ and\ \citenamefont {Granick}}]{Jiang2014}%
  \BibitemOpen
  \bibfield  {author} {\bibinfo {author} {\bibfnamefont {S.}~\bibnamefont {Jiang}}, \bibinfo {author} {\bibfnamefont {J.}~\bibnamefont {Yan}}, \bibinfo {author} {\bibfnamefont {J.~K.}\ \bibnamefont {Whitmer}}, \bibinfo {author} {\bibfnamefont {S.~M.}\ \bibnamefont {Anthony}}, \bibinfo {author} {\bibfnamefont {E.}~\bibnamefont {Luijten}},\ and\ \bibinfo {author} {\bibfnamefont {S.}~\bibnamefont {Granick}},\ }\bibfield  {title} {\bibinfo {title} {Orientationally glassy crystals of {J}anus spheres},\ }\href {https://doi.org/10.1103/PhysRevLett.112.218301} {\bibfield  {journal} {\bibinfo  {journal} {Phys. Rev. Lett.}\ }\textbf {\bibinfo {volume} {112}},\ \bibinfo {pages} {218301} (\bibinfo {year} {2014})}\BibitemShut {NoStop}%
\bibitem [{\citenamefont {Mitsumoto}\ and\ \citenamefont {Yoshino}(2018)}]{Mitsumoto2018}%
  \BibitemOpen
  \bibfield  {author} {\bibinfo {author} {\bibfnamefont {K.}~\bibnamefont {Mitsumoto}}\ and\ \bibinfo {author} {\bibfnamefont {H.}~\bibnamefont {Yoshino}},\ }\bibfield  {title} {\bibinfo {title} {Orientational ordering of closely packed {J}anus particles},\ }\href {https://doi.org/10.1039/C8SM00622A} {\bibfield  {journal} {\bibinfo  {journal} {Soft Matter}\ }\textbf {\bibinfo {volume} {14}},\ \bibinfo {pages} {3919} (\bibinfo {year} {2018})}\BibitemShut {NoStop}%
\bibitem [{\citenamefont {Huang}\ \emph {et~al.}(2019)\citenamefont {Huang}, \citenamefont {Zhu}, \citenamefont {Chen}, \citenamefont {Hou},\ and\ \citenamefont {Yan}}]{Huang2019}%
  \BibitemOpen
  \bibfield  {author} {\bibinfo {author} {\bibfnamefont {Z.}~\bibnamefont {Huang}}, \bibinfo {author} {\bibfnamefont {G.}~\bibnamefont {Zhu}}, \bibinfo {author} {\bibfnamefont {P.}~\bibnamefont {Chen}}, \bibinfo {author} {\bibfnamefont {C.}~\bibnamefont {Hou}},\ and\ \bibinfo {author} {\bibfnamefont {L.-T.}\ \bibnamefont {Yan}},\ }\bibfield  {title} {\bibinfo {title} {Plastic crystal-to-crystal transition of {J}anus particles under shear},\ }\href {https://doi.org/10.1103/PhysRevLett.122.198002} {\bibfield  {journal} {\bibinfo  {journal} {Phys. Rev. Lett.}\ }\textbf {\bibinfo {volume} {122}},\ \bibinfo {pages} {198002} (\bibinfo {year} {2019})}\BibitemShut {NoStop}%
\bibitem [{\citenamefont {Patrykiejew}\ and\ \citenamefont {Rżysko}(2021)}]{Patrykiejew2021}%
  \BibitemOpen
  \bibfield  {author} {\bibinfo {author} {\bibfnamefont {A.}~\bibnamefont {Patrykiejew}}\ and\ \bibinfo {author} {\bibfnamefont {W.}~\bibnamefont {Rżysko}},\ }\bibfield  {title} {\bibinfo {title} {The order–disorder transitions in systems of {J}anus-like particles on a triangular lattice-revisited},\ }\href {https://doi.org/https://doi.org/10.1016/j.physa.2021.125819} {\bibfield  {journal} {\bibinfo  {journal} {Physica A}\ }\textbf {\bibinfo {volume} {570}},\ \bibinfo {pages} {125819} (\bibinfo {year} {2021})}\BibitemShut {NoStop}%
\bibitem [{\citenamefont {Liang}\ \emph {et~al.}(2021)\citenamefont {Liang}, \citenamefont {Ma},\ and\ \citenamefont {Olvera de~la Cruz}}]{Liang2021}%
  \BibitemOpen
  \bibfield  {author} {\bibinfo {author} {\bibfnamefont {Y.}~\bibnamefont {Liang}}, \bibinfo {author} {\bibfnamefont {B.}~\bibnamefont {Ma}},\ and\ \bibinfo {author} {\bibfnamefont {M.}~\bibnamefont {Olvera de~la Cruz}},\ }\bibfield  {title} {\bibinfo {title} {Reverse order-disorder transition of {J}anus particles confined in two dimensions},\ }\href {https://doi.org/10.1103/PhysRevE.103.062607} {\bibfield  {journal} {\bibinfo  {journal} {Phys. Rev. E}\ }\textbf {\bibinfo {volume} {103}},\ \bibinfo {pages} {062607} (\bibinfo {year} {2021})}\BibitemShut {NoStop}%
\bibitem [{\citenamefont {Hu}\ \emph {et~al.}(2022)\citenamefont {Hu}, \citenamefont {Ziff},\ and\ \citenamefont {Deng}}]{Hu2022}%
  \BibitemOpen
  \bibfield  {author} {\bibinfo {author} {\bibfnamefont {H.}~\bibnamefont {Hu}}, \bibinfo {author} {\bibfnamefont {R.~M.}\ \bibnamefont {Ziff}},\ and\ \bibinfo {author} {\bibfnamefont {Y.}~\bibnamefont {Deng}},\ }\bibfield  {title} {\bibinfo {title} {Universal critical behavior of percolation in orientationally ordered {J}anus particles and other anisotropic systems},\ }\href {https://doi.org/10.1103/PhysRevLett.129.278002} {\bibfield  {journal} {\bibinfo  {journal} {Phys. Rev. Lett.}\ }\textbf {\bibinfo {volume} {129}},\ \bibinfo {pages} {278002} (\bibinfo {year} {2022})}\BibitemShut {NoStop}%
\bibitem [{\citenamefont {Huang}\ \emph {et~al.}(2022)\citenamefont {Huang}, \citenamefont {Zeng}, \citenamefont {Wang}, \citenamefont {Chen},\ and\ \citenamefont {Han}}]{Huang2022}%
  \BibitemOpen
  \bibfield  {author} {\bibinfo {author} {\bibfnamefont {T.}~\bibnamefont {Huang}}, \bibinfo {author} {\bibfnamefont {C.}~\bibnamefont {Zeng}}, \bibinfo {author} {\bibfnamefont {H.}~\bibnamefont {Wang}}, \bibinfo {author} {\bibfnamefont {Y.}~\bibnamefont {Chen}},\ and\ \bibinfo {author} {\bibfnamefont {Y.}~\bibnamefont {Han}},\ }\bibfield  {title} {\bibinfo {title} {{Internal-stress-induced solid-solid transition involving orientational domains of anisotropic particles}},\ }\href {https://doi.org/10.1103/PhysRevE.106.014612} {\bibfield  {journal} {\bibinfo  {journal} {Phys. Rev. E}\ }\textbf {\bibinfo {volume} {106}},\ \bibinfo {pages} {014612} (\bibinfo {year} {2022})}\BibitemShut {NoStop}%
\bibitem [{\citenamefont {Chen}\ \emph {et~al.}(2011)\citenamefont {Chen}, \citenamefont {Bae},\ and\ \citenamefont {Granick}}]{ChenGranick2011}%
  \BibitemOpen
  \bibfield  {author} {\bibinfo {author} {\bibfnamefont {Q.}~\bibnamefont {Chen}}, \bibinfo {author} {\bibfnamefont {S.~C.}\ \bibnamefont {Bae}},\ and\ \bibinfo {author} {\bibfnamefont {S.}~\bibnamefont {Granick}},\ }\bibfield  {title} {\bibinfo {title} {Directed self-assembly of a colloidal kagome lattice},\ }\href {https://doi.org/https://doi.org/10.1038/nature09713} {\bibfield  {journal} {\bibinfo  {journal} {Nature}\ }\textbf {\bibinfo {volume} {469}},\ \bibinfo {pages} {381} (\bibinfo {year} {2011})}\BibitemShut {NoStop}%
\bibitem [{\citenamefont {Romano}\ and\ \citenamefont {Sciortino}(2011)}]{RomanoSciortino2011}%
  \BibitemOpen
  \bibfield  {author} {\bibinfo {author} {\bibfnamefont {F.}~\bibnamefont {Romano}}\ and\ \bibinfo {author} {\bibfnamefont {F.}~\bibnamefont {Sciortino}},\ }\bibfield  {title} {\bibinfo {title} {Two dimensional assembly of triblock {Janus} particles into crystal phases in the two bond per patch limit},\ }\href {https://doi.org/10.1039/C0SM01494J} {\bibfield  {journal} {\bibinfo  {journal} {Soft Matter}\ }\textbf {\bibinfo {volume} {7}},\ \bibinfo {pages} {5799} (\bibinfo {year} {2011})}\BibitemShut {NoStop}%
\bibitem [{\citenamefont {Eslami}\ \emph {et~al.}(2018)\citenamefont {Eslami}, \citenamefont {Bahri},\ and\ \citenamefont {Müller-Plathe}}]{Eslami2018}%
  \BibitemOpen
  \bibfield  {author} {\bibinfo {author} {\bibfnamefont {H.}~\bibnamefont {Eslami}}, \bibinfo {author} {\bibfnamefont {K.}~\bibnamefont {Bahri}},\ and\ \bibinfo {author} {\bibfnamefont {F.}~\bibnamefont {Müller-Plathe}},\ }\bibfield  {title} {\bibinfo {title} {Solid–liquid and solid–solid phase diagrams of self-assembled triblock {J}anus nanoparticles from solution},\ }\href {https://doi.org/10.1021/acs.jpcc.8b02043} {\bibfield  {journal} {\bibinfo  {journal} {J. Phys. Chem. C}\ }\textbf {\bibinfo {volume} {122}},\ \bibinfo {pages} {9235} (\bibinfo {year} {2018})}\BibitemShut {NoStop}%
\bibitem [{\citenamefont {Mallory}\ and\ \citenamefont {Cacciuto}(2019)}]{Mallory2019}%
  \BibitemOpen
  \bibfield  {author} {\bibinfo {author} {\bibfnamefont {S.~A.}\ \bibnamefont {Mallory}}\ and\ \bibinfo {author} {\bibfnamefont {A.}~\bibnamefont {Cacciuto}},\ }\bibfield  {title} {\bibinfo {title} {Activity-enhanced self-assembly of a colloidal kagome lattice},\ }\href {https://doi.org/10.1021/jacs.8b12165} {\bibfield  {journal} {\bibinfo  {journal} {J. Am. Chem. Soc.}\ }\textbf {\bibinfo {volume} {141}},\ \bibinfo {pages} {2500} (\bibinfo {year} {2019})}\BibitemShut {NoStop}%
\bibitem [{\citenamefont {Bahri}\ \emph {et~al.}(2022)\citenamefont {Bahri}, \citenamefont {Eslami},\ and\ \citenamefont {Müller-Plathe}}]{Bahri2022}%
  \BibitemOpen
  \bibfield  {author} {\bibinfo {author} {\bibfnamefont {K.}~\bibnamefont {Bahri}}, \bibinfo {author} {\bibfnamefont {H.}~\bibnamefont {Eslami}},\ and\ \bibinfo {author} {\bibfnamefont {F.}~\bibnamefont {Müller-Plathe}},\ }\bibfield  {title} {\bibinfo {title} {Self-assembly of model triblock {J}anus colloidal particles in two dimensions},\ }\href {https://doi.org/10.1021/acs.jctc.1c01116} {\bibfield  {journal} {\bibinfo  {journal} {J. Chem. Theory Comput.}\ }\textbf {\bibinfo {volume} {18}},\ \bibinfo {pages} {1870} (\bibinfo {year} {2022})}\BibitemShut {NoStop}%
\bibitem [{\citenamefont {Schubert}\ \emph {et~al.}(2025)\citenamefont {Schubert}, \citenamefont {Navas},\ and\ \citenamefont {Klapp}}]{Schubert2025}%
  \BibitemOpen
  \bibfield  {author} {\bibinfo {author} {\bibfnamefont {J.~F.}\ \bibnamefont {Schubert}}, \bibinfo {author} {\bibfnamefont {S.~F.}\ \bibnamefont {Navas}},\ and\ \bibinfo {author} {\bibfnamefont {S.~H.~L.}\ \bibnamefont {Klapp}},\ }\href {https://arxiv.org/abs/2504.20764} {\bibinfo {title} {Self-assembly and time-dependent control of active and passive triblock {J}anus colloids}} (\bibinfo {year} {2025}),\ \Eprint {https://arxiv.org/abs/2504.20764} {arXiv:2504.20764 [cond-mat.soft]} \BibitemShut {NoStop}%
\bibitem [{\citenamefont {Mao}\ \emph {et~al.}(2013)\citenamefont {Mao}, \citenamefont {Chen},\ and\ \citenamefont {Granick}}]{Mao2013}%
  \BibitemOpen
  \bibfield  {author} {\bibinfo {author} {\bibfnamefont {X.}~\bibnamefont {Mao}}, \bibinfo {author} {\bibfnamefont {Q.}~\bibnamefont {Chen}},\ and\ \bibinfo {author} {\bibfnamefont {S.}~\bibnamefont {Granick}},\ }\bibfield  {title} {\bibinfo {title} {Entropy favours open colloidal lattices},\ }\href {https://doi.org/10.1038/nmat3496} {\bibfield  {journal} {\bibinfo  {journal} {Nature Materials}\ }\textbf {\bibinfo {volume} {12}},\ \bibinfo {pages} {217} (\bibinfo {year} {2013})}\BibitemShut {NoStop}%
\bibitem [{\citenamefont {Nussinov}\ and\ \citenamefont {van~den Brink}(2015)}]{Nussinov2015}%
  \BibitemOpen
  \bibfield  {author} {\bibinfo {author} {\bibfnamefont {Z.}~\bibnamefont {Nussinov}}\ and\ \bibinfo {author} {\bibfnamefont {J.}~\bibnamefont {van~den Brink}},\ }\bibfield  {title} {\bibinfo {title} {Compass models: Theory and physical motivations},\ }\href {https://doi.org/10.1103/RevModPhys.87.1} {\bibfield  {journal} {\bibinfo  {journal} {Rev. Mod. Phys.}\ }\textbf {\bibinfo {volume} {87}},\ \bibinfo {pages} {1} (\bibinfo {year} {2015})}\BibitemShut {NoStop}%
\bibitem [{\citenamefont {Placke}\ \emph {et~al.}(2023)\citenamefont {Placke}, \citenamefont {Sommers}, \citenamefont {Sondhi},\ and\ \citenamefont {Moessner}}]{Placke2023}%
  \BibitemOpen
  \bibfield  {author} {\bibinfo {author} {\bibfnamefont {B.}~\bibnamefont {Placke}}, \bibinfo {author} {\bibfnamefont {G.~M.}\ \bibnamefont {Sommers}}, \bibinfo {author} {\bibfnamefont {S.~L.}\ \bibnamefont {Sondhi}},\ and\ \bibinfo {author} {\bibfnamefont {R.}~\bibnamefont {Moessner}},\ }\bibfield  {title} {\bibinfo {title} {Arresting dynamics in hardcore spin models},\ }\href {https://doi.org/10.1103/PhysRevB.107.L180302} {\bibfield  {journal} {\bibinfo  {journal} {Phys. Rev. B}\ }\textbf {\bibinfo {volume} {107}},\ \bibinfo {pages} {L180302} (\bibinfo {year} {2023})}\BibitemShut {NoStop}%
\bibitem [{\citenamefont {Saryal}\ and\ \citenamefont {Dhar}(2023)}]{Saryal2023}%
  \BibitemOpen
  \bibfield  {author} {\bibinfo {author} {\bibfnamefont {S.}~\bibnamefont {Saryal}}\ and\ \bibinfo {author} {\bibfnamefont {D.}~\bibnamefont {Dhar}},\ }\bibfield  {title} {\bibinfo {title} {Cusp singularities in the distribution of orientations of asymmetrically pivoted hard disks on a lattice},\ }\href {https://doi.org/10.1103/PhysRevE.108.044110} {\bibfield  {journal} {\bibinfo  {journal} {Phys. Rev. E}\ }\textbf {\bibinfo {volume} {108}},\ \bibinfo {pages} {044110} (\bibinfo {year} {2023})}\BibitemShut {NoStop}%
\bibitem [{\citenamefont {Dadhichi}\ \emph {et~al.}(2020)\citenamefont {Dadhichi}, \citenamefont {Kethapelli}, \citenamefont {Chajwa}, \citenamefont {Ramaswamy},\ and\ \citenamefont {Maitra}}]{Dadhichi2020}%
  \BibitemOpen
  \bibfield  {author} {\bibinfo {author} {\bibfnamefont {L.~P.}\ \bibnamefont {Dadhichi}}, \bibinfo {author} {\bibfnamefont {J.}~\bibnamefont {Kethapelli}}, \bibinfo {author} {\bibfnamefont {R.}~\bibnamefont {Chajwa}}, \bibinfo {author} {\bibfnamefont {S.}~\bibnamefont {Ramaswamy}},\ and\ \bibinfo {author} {\bibfnamefont {A.}~\bibnamefont {Maitra}},\ }\bibfield  {title} {\bibinfo {title} {Nonmutual torques and the unimportance of motility for long-range order in two-dimensional flocks},\ }\href {https://doi.org/10.1103/PhysRevE.101.052601} {\bibfield  {journal} {\bibinfo  {journal} {Phys. Rev. E}\ }\textbf {\bibinfo {volume} {101}},\ \bibinfo {pages} {052601} (\bibinfo {year} {2020})}\BibitemShut {NoStop}%
\bibitem [{\citenamefont {Dopierala}\ \emph {et~al.}(2025)\citenamefont {Dopierala}, \citenamefont {Chat\'e}, \citenamefont {Shi},\ and\ \citenamefont {Solon}}]{Dopierala2025}%
  \BibitemOpen
  \bibfield  {author} {\bibinfo {author} {\bibfnamefont {D.}~\bibnamefont {Dopierala}}, \bibinfo {author} {\bibfnamefont {H.}~\bibnamefont {Chat\'e}}, \bibinfo {author} {\bibfnamefont {X.-q.}\ \bibnamefont {Shi}},\ and\ \bibinfo {author} {\bibfnamefont {A.}~\bibnamefont {Solon}},\ }\bibfield  {title} {\bibinfo {title} {Inescapable anisotropy of nonreciprocal {XY} models},\ }\href {https://doi.org/10.1103/r3dx-7lrd} {\bibfield  {journal} {\bibinfo  {journal} {Phys. Rev. Lett.}\ }\textbf {\bibinfo {volume} {135}},\ \bibinfo {pages} {088302} (\bibinfo {year} {2025})}\BibitemShut {NoStop}%
\bibitem [{\citenamefont {Popli}\ \emph {et~al.}(2025)\citenamefont {Popli}, \citenamefont {Maitra},\ and\ \citenamefont {Ramaswamy}}]{Popli2025}%
  \BibitemOpen
  \bibfield  {author} {\bibinfo {author} {\bibfnamefont {P.}~\bibnamefont {Popli}}, \bibinfo {author} {\bibfnamefont {A.}~\bibnamefont {Maitra}},\ and\ \bibinfo {author} {\bibfnamefont {S.}~\bibnamefont {Ramaswamy}},\ }\bibfield  {title} {\bibinfo {title} {Ordering and defect cloaking in nonreciprocal lattice {XY} models},\ }\href {https://doi.org/10.1103/2yky-45sr} {\bibfield  {journal} {\bibinfo  {journal} {Phys. Rev. Lett.}\ }\textbf {\bibinfo {volume} {135}},\ \bibinfo {pages} {088303} (\bibinfo {year} {2025})}\BibitemShut {NoStop}%
\bibitem [{\citenamefont {Loos}\ \emph {et~al.}(2023)\citenamefont {Loos}, \citenamefont {Klapp},\ and\ \citenamefont {Martynec}}]{Loos2023}%
  \BibitemOpen
  \bibfield  {author} {\bibinfo {author} {\bibfnamefont {S.~A.~M.}\ \bibnamefont {Loos}}, \bibinfo {author} {\bibfnamefont {S.~H.~L.}\ \bibnamefont {Klapp}},\ and\ \bibinfo {author} {\bibfnamefont {T.}~\bibnamefont {Martynec}},\ }\bibfield  {title} {\bibinfo {title} {Long-range order and directional defect propagation in the nonreciprocal {XY} model with vision cone interactions},\ }\href {https://doi.org/10.1103/PhysRevLett.130.198301} {\bibfield  {journal} {\bibinfo  {journal} {Phys. Rev. Lett.}\ }\textbf {\bibinfo {volume} {130}},\ \bibinfo {pages} {198301} (\bibinfo {year} {2023})}\BibitemShut {NoStop}%
\bibitem [{\citenamefont {Bandini}\ \emph {et~al.}(2025)\citenamefont {Bandini}, \citenamefont {Venturelli}, \citenamefont {Loos}, \citenamefont {Jelic},\ and\ \citenamefont {Gambassi}}]{Bandini2024}%
  \BibitemOpen
  \bibfield  {author} {\bibinfo {author} {\bibfnamefont {G.}~\bibnamefont {Bandini}}, \bibinfo {author} {\bibfnamefont {D.}~\bibnamefont {Venturelli}}, \bibinfo {author} {\bibfnamefont {S.~A.~M.}\ \bibnamefont {Loos}}, \bibinfo {author} {\bibfnamefont {A.}~\bibnamefont {Jelic}},\ and\ \bibinfo {author} {\bibfnamefont {A.}~\bibnamefont {Gambassi}},\ }\bibfield  {title} {\bibinfo {title} {The {XY} model with vision cone: non-reciprocal vs. reciprocal interactions},\ }\href {https://doi.org/10.1088/1742-5468/adc243} {\bibfield  {journal} {\bibinfo  {journal} {J. Stat. Mech.}\ ,\ \bibinfo {pages} {053205}} (\bibinfo {year} {2025})}\BibitemShut {NoStop}%
\bibitem [{\citenamefont {Liu}\ \emph {et~al.}(2025)\citenamefont {Liu}, \citenamefont {Zheng}, \citenamefont {Nian},\ and\ \citenamefont {Xiong}}]{Liu2025}%
  \BibitemOpen
  \bibfield  {author} {\bibinfo {author} {\bibfnamefont {Z.-Y.}\ \bibnamefont {Liu}}, \bibinfo {author} {\bibfnamefont {B.}~\bibnamefont {Zheng}}, \bibinfo {author} {\bibfnamefont {L.-L.}\ \bibnamefont {Nian}},\ and\ \bibinfo {author} {\bibfnamefont {L.}~\bibnamefont {Xiong}},\ }\bibfield  {title} {\bibinfo {title} {Dynamic approach to the two-dimensional nonreciprocal {XY} model with vision cone interactions},\ }\href {https://doi.org/10.1103/PhysRevE.111.014131} {\bibfield  {journal} {\bibinfo  {journal} {Phys. Rev. E}\ }\textbf {\bibinfo {volume} {111}},\ \bibinfo {pages} {014131} (\bibinfo {year} {2025})}\BibitemShut {NoStop}%
\bibitem [{\citenamefont {Jos\'e}\ \emph {et~al.}(1977)\citenamefont {Jos\'e}, \citenamefont {Kadanoff}, \citenamefont {Kirkpatrick},\ and\ \citenamefont {Nelson}}]{Jose1977}%
  \BibitemOpen
  \bibfield  {author} {\bibinfo {author} {\bibfnamefont {J.~V.}\ \bibnamefont {Jos\'e}}, \bibinfo {author} {\bibfnamefont {L.~P.}\ \bibnamefont {Kadanoff}}, \bibinfo {author} {\bibfnamefont {S.}~\bibnamefont {Kirkpatrick}},\ and\ \bibinfo {author} {\bibfnamefont {D.~R.}\ \bibnamefont {Nelson}},\ }\bibfield  {title} {\bibinfo {title} {Renormalization, vortices, and symmetry-breaking perturbations in the two-dimensional planar model},\ }\href {https://doi.org/10.1103/PhysRevB.16.1217} {\bibfield  {journal} {\bibinfo  {journal} {Phys. Rev. B}\ }\textbf {\bibinfo {volume} {16}},\ \bibinfo {pages} {1217} (\bibinfo {year} {1977})}\BibitemShut {NoStop}%
\bibitem [{\citenamefont {Elitzur}\ \emph {et~al.}(1979)\citenamefont {Elitzur}, \citenamefont {Pearson},\ and\ \citenamefont {Shigemitsu}}]{Elitzur1979}%
  \BibitemOpen
  \bibfield  {author} {\bibinfo {author} {\bibfnamefont {S.}~\bibnamefont {Elitzur}}, \bibinfo {author} {\bibfnamefont {R.~B.}\ \bibnamefont {Pearson}},\ and\ \bibinfo {author} {\bibfnamefont {J.}~\bibnamefont {Shigemitsu}},\ }\bibfield  {title} {\bibinfo {title} {Phase structure of discrete {A}belian spin and gauge systems},\ }\href {https://doi.org/10.1103/PhysRevD.19.3698} {\bibfield  {journal} {\bibinfo  {journal} {Phys. Rev. D}\ }\textbf {\bibinfo {volume} {19}},\ \bibinfo {pages} {3698} (\bibinfo {year} {1979})}\BibitemShut {NoStop}%
\bibitem [{\citenamefont {Tomita}\ and\ \citenamefont {Okabe}(2002)}]{Tomita2002}%
  \BibitemOpen
  \bibfield  {author} {\bibinfo {author} {\bibfnamefont {Y.}~\bibnamefont {Tomita}}\ and\ \bibinfo {author} {\bibfnamefont {Y.}~\bibnamefont {Okabe}},\ }\bibfield  {title} {\bibinfo {title} {Probability-changing cluster algorithm for two-dimensional {XY} and clock models},\ }\href {https://doi.org/10.1103/PhysRevB.65.184405} {\bibfield  {journal} {\bibinfo  {journal} {Phys. Rev. B}\ }\textbf {\bibinfo {volume} {65}},\ \bibinfo {pages} {184405} (\bibinfo {year} {2002})}\BibitemShut {NoStop}%
\bibitem [{\citenamefont {Ueda}\ \emph {et~al.}(2020)\citenamefont {Ueda}, \citenamefont {Okunishi}, \citenamefont {Harada}, \citenamefont {Kr\ifmmode~\check{c}\else \v{c}\fi{}m\'ar}, \citenamefont {Gendiar}, \citenamefont {Yunoki},\ and\ \citenamefont {Nishino}}]{Ueda2020}%
  \BibitemOpen
  \bibfield  {author} {\bibinfo {author} {\bibfnamefont {H.}~\bibnamefont {Ueda}}, \bibinfo {author} {\bibfnamefont {K.}~\bibnamefont {Okunishi}}, \bibinfo {author} {\bibfnamefont {K.}~\bibnamefont {Harada}}, \bibinfo {author} {\bibfnamefont {R.}~\bibnamefont {Kr\ifmmode~\check{c}\else \v{c}\fi{}m\'ar}}, \bibinfo {author} {\bibfnamefont {A.}~\bibnamefont {Gendiar}}, \bibinfo {author} {\bibfnamefont {S.}~\bibnamefont {Yunoki}},\ and\ \bibinfo {author} {\bibfnamefont {T.}~\bibnamefont {Nishino}},\ }\bibfield  {title} {\bibinfo {title} {Finite-$m$ scaling analysis of {B}erezinskii-{K}osterlitz-{T}houless phase transitions and entanglement spectrum for the six-state clock model},\ }\href {https://doi.org/10.1103/PhysRevE.101.062111} {\bibfield  {journal} {\bibinfo  {journal} {Phys. Rev. E}\ }\textbf {\bibinfo {volume} {101}},\ \bibinfo {pages} {062111} (\bibinfo {year} {2020})}\BibitemShut {NoStop}%
\bibitem [{\citenamefont {Chen}\ \emph {et~al.}(2022)\citenamefont {Chen}, \citenamefont {Hou}, \citenamefont {Fang},\ and\ \citenamefont {Deng}}]{Chen2022}%
  \BibitemOpen
  \bibfield  {author} {\bibinfo {author} {\bibfnamefont {H.}~\bibnamefont {Chen}}, \bibinfo {author} {\bibfnamefont {P.}~\bibnamefont {Hou}}, \bibinfo {author} {\bibfnamefont {S.}~\bibnamefont {Fang}},\ and\ \bibinfo {author} {\bibfnamefont {Y.}~\bibnamefont {Deng}},\ }\bibfield  {title} {\bibinfo {title} {Monte carlo study of duality and the {B}erezinskii-{K}osterlitz-{T}houless phase transitions of the two-dimensional $q$-state clock model in flow representations},\ }\href {https://doi.org/10.1103/PhysRevE.106.024106} {\bibfield  {journal} {\bibinfo  {journal} {Phys. Rev. E}\ }\textbf {\bibinfo {volume} {106}},\ \bibinfo {pages} {024106} (\bibinfo {year} {2022})}\BibitemShut {NoStop}%
\bibitem [{\citenamefont {Tuan}\ \emph {et~al.}(2022)\citenamefont {Tuan}, \citenamefont {Long}, \citenamefont {Nui}, \citenamefont {Minh}, \citenamefont {Trung~Kien},\ and\ \citenamefont {Viet}}]{Tuan2022}%
  \BibitemOpen
  \bibfield  {author} {\bibinfo {author} {\bibfnamefont {L.~M.}\ \bibnamefont {Tuan}}, \bibinfo {author} {\bibfnamefont {T.~T.}\ \bibnamefont {Long}}, \bibinfo {author} {\bibfnamefont {D.~X.}\ \bibnamefont {Nui}}, \bibinfo {author} {\bibfnamefont {P.~T.}\ \bibnamefont {Minh}}, \bibinfo {author} {\bibfnamefont {N.~D.}\ \bibnamefont {Trung~Kien}},\ and\ \bibinfo {author} {\bibfnamefont {D.~X.}\ \bibnamefont {Viet}},\ }\bibfield  {title} {\bibinfo {title} {Binder ratio in the two-dimensional $q$-state clock model},\ }\href {https://doi.org/10.1103/PhysRevE.106.034138} {\bibfield  {journal} {\bibinfo  {journal} {Phys. Rev. E}\ }\textbf {\bibinfo {volume} {106}},\ \bibinfo {pages} {034138} (\bibinfo {year} {2022})}\BibitemShut {NoStop}%
\bibitem [{\citenamefont {Bianchi}\ \emph {et~al.}(2006)\citenamefont {Bianchi}, \citenamefont {Largo}, \citenamefont {Tartaglia}, \citenamefont {Zaccarelli},\ and\ \citenamefont {Sciortino}}]{Bianchi2006}%
  \BibitemOpen
  \bibfield  {author} {\bibinfo {author} {\bibfnamefont {E.}~\bibnamefont {Bianchi}}, \bibinfo {author} {\bibfnamefont {J.}~\bibnamefont {Largo}}, \bibinfo {author} {\bibfnamefont {P.}~\bibnamefont {Tartaglia}}, \bibinfo {author} {\bibfnamefont {E.}~\bibnamefont {Zaccarelli}},\ and\ \bibinfo {author} {\bibfnamefont {F.}~\bibnamefont {Sciortino}},\ }\bibfield  {title} {\bibinfo {title} {Phase diagram of patchy colloids: Towards empty liquids},\ }\href {https://doi.org/10.1103/PhysRevLett.97.168301} {\bibfield  {journal} {\bibinfo  {journal} {Phys. Rev. Lett.}\ }\textbf {\bibinfo {volume} {97}},\ \bibinfo {pages} {168301} (\bibinfo {year} {2006})}\BibitemShut {NoStop}%
\bibitem [{\citenamefont {Tavares}\ \emph {et~al.}(2009)\citenamefont {Tavares}, \citenamefont {Teixeira},\ and\ \citenamefont {Telo~da Gama}}]{Tavares2009}%
  \BibitemOpen
  \bibfield  {author} {\bibinfo {author} {\bibfnamefont {J.~M.}\ \bibnamefont {Tavares}}, \bibinfo {author} {\bibfnamefont {P.~I.~C.}\ \bibnamefont {Teixeira}},\ and\ \bibinfo {author} {\bibfnamefont {M.~M.}\ \bibnamefont {Telo~da Gama}},\ }\bibfield  {title} {\bibinfo {title} {Criticality of colloids with distinct interaction patches: The limits of linear chains, hyperbranched polymers, and dimers},\ }\href {https://doi.org/10.1103/PhysRevE.80.021506} {\bibfield  {journal} {\bibinfo  {journal} {Phys. Rev. E}\ }\textbf {\bibinfo {volume} {80}},\ \bibinfo {pages} {021506} (\bibinfo {year} {2009})}\BibitemShut {NoStop}%
\bibitem [{\citenamefont {Swinkels}\ \emph {et~al.}(2024)\citenamefont {Swinkels}, \citenamefont {Sinaasappel}, \citenamefont {Gong}, \citenamefont {Sacanna}, \citenamefont {Meyer}, \citenamefont {Sciortino},\ and\ \citenamefont {Schall}}]{Swinkels2024}%
  \BibitemOpen
  \bibfield  {author} {\bibinfo {author} {\bibfnamefont {P.~J.~M.}\ \bibnamefont {Swinkels}}, \bibinfo {author} {\bibfnamefont {R.}~\bibnamefont {Sinaasappel}}, \bibinfo {author} {\bibfnamefont {Z.}~\bibnamefont {Gong}}, \bibinfo {author} {\bibfnamefont {S.}~\bibnamefont {Sacanna}}, \bibinfo {author} {\bibfnamefont {W.~V.}\ \bibnamefont {Meyer}}, \bibinfo {author} {\bibfnamefont {F.}~\bibnamefont {Sciortino}},\ and\ \bibinfo {author} {\bibfnamefont {P.}~\bibnamefont {Schall}},\ }\bibfield  {title} {\bibinfo {title} {Networks of limited-valency patchy particles},\ }\href {https://doi.org/10.1103/PhysRevLett.132.078203} {\bibfield  {journal} {\bibinfo  {journal} {Phys. Rev. Lett.}\ }\textbf {\bibinfo {volume} {132}},\ \bibinfo {pages} {078203} (\bibinfo {year} {2024})}\BibitemShut {NoStop}%
\bibitem [{\citenamefont {Wang}\ \emph {et~al.}(2022)\citenamefont {Wang}, \citenamefont {He}, \citenamefont {Wang},\ and\ \citenamefont {Hu}}]{WangHu2022}%
  \BibitemOpen
  \bibfield  {author} {\bibinfo {author} {\bibfnamefont {Q.}~\bibnamefont {Wang}}, \bibinfo {author} {\bibfnamefont {Z.}~\bibnamefont {He}}, \bibinfo {author} {\bibfnamefont {J.}~\bibnamefont {Wang}},\ and\ \bibinfo {author} {\bibfnamefont {H.}~\bibnamefont {Hu}},\ }\bibfield  {title} {\bibinfo {title} {Percolation thresholds of randomly rotating patchy particles on archimedean lattices},\ }\href {https://doi.org/10.1103/PhysRevE.105.034118} {\bibfield  {journal} {\bibinfo  {journal} {Phys. Rev. E}\ }\textbf {\bibinfo {volume} {105}},\ \bibinfo {pages} {034118} (\bibinfo {year} {2022})}\BibitemShut {NoStop}%
\bibitem [{\citenamefont {Wang}\ \emph {et~al.}(2024)\citenamefont {Wang}, \citenamefont {Wang}, \citenamefont {Liu},\ and\ \citenamefont {Hu}}]{WangHu2024}%
  \BibitemOpen
  \bibfield  {author} {\bibinfo {author} {\bibfnamefont {J.}~\bibnamefont {Wang}}, \bibinfo {author} {\bibfnamefont {X.}~\bibnamefont {Wang}}, \bibinfo {author} {\bibfnamefont {W.}~\bibnamefont {Liu}},\ and\ \bibinfo {author} {\bibfnamefont {H.}~\bibnamefont {Hu}},\ }\bibfield  {title} {\bibinfo {title} {Percolation thresholds of disks with random nonoverlapping patches on four regular two-dimensional lattices},\ }\href {https://doi.org/10.1103/PhysRevE.109.064104} {\bibfield  {journal} {\bibinfo  {journal} {Phys. Rev. E}\ }\textbf {\bibinfo {volume} {109}},\ \bibinfo {pages} {064104} (\bibinfo {year} {2024})}\BibitemShut {NoStop}%
\bibitem [{\citenamefont {Barber}(1983)}]{Barber1983}%
  \BibitemOpen
  \bibfield  {author} {\bibinfo {author} {\bibfnamefont {M.~N.}\ \bibnamefont {Barber}},\ }\href@noop {} {\emph {\bibinfo {title} {``Finite-size scaling", {\rm in {P}hase {T}ransitions and {C}ritical {P}henomena}}}},\ edited by\ \bibinfo {editor} {\bibfnamefont {C.}~\bibnamefont {Domb}}\ and\ \bibinfo {editor} {\bibfnamefont {J.~L.}\ \bibnamefont {Lebowitz}},\ Vol.~\bibinfo {volume} {8}\ (\bibinfo  {publisher} {Academic Press},\ \bibinfo {address} {New York},\ \bibinfo {year} {1983})\BibitemShut {NoStop}%
\bibitem [{\citenamefont {Chen}\ \emph {et~al.}(2019)\citenamefont {Chen}, \citenamefont {Hu}, \citenamefont {Izmailian},\ and\ \citenamefont {Wu}}]{Chen2019}%
  \BibitemOpen
  \bibfield  {author} {\bibinfo {author} {\bibfnamefont {C.-N.}\ \bibnamefont {Chen}}, \bibinfo {author} {\bibfnamefont {C.-K.}\ \bibnamefont {Hu}}, \bibinfo {author} {\bibfnamefont {N.~S.}\ \bibnamefont {Izmailian}},\ and\ \bibinfo {author} {\bibfnamefont {M.-C.}\ \bibnamefont {Wu}},\ }\bibfield  {title} {\bibinfo {title} {Specific heat and partition function zeros for the dimer model on the checkerboard {$B$} lattice: Finite-size effects},\ }\href {https://doi.org/10.1103/PhysRevE.99.012102} {\bibfield  {journal} {\bibinfo  {journal} {Phys. Rev. E}\ }\textbf {\bibinfo {volume} {99}},\ \bibinfo {pages} {012102} (\bibinfo {year} {2019})}\BibitemShut {NoStop}%
\bibitem [{\citenamefont {dos Santos}\ \emph {et~al.}(2023)\citenamefont {dos Santos}, \citenamefont {Cisternas}, \citenamefont {Vogel},\ and\ \citenamefont {Ramirez-Pastor}}]{dosSantos2023}%
  \BibitemOpen
  \bibfield  {author} {\bibinfo {author} {\bibfnamefont {G.}~\bibnamefont {dos Santos}}, \bibinfo {author} {\bibfnamefont {E.}~\bibnamefont {Cisternas}}, \bibinfo {author} {\bibfnamefont {E.~E.}\ \bibnamefont {Vogel}},\ and\ \bibinfo {author} {\bibfnamefont {A.~J.}\ \bibnamefont {Ramirez-Pastor}},\ }\bibfield  {title} {\bibinfo {title} {Orientational phase transition in monolayers of multipolar straight rigid rods: The case of 2-thiophene molecule adsorption on the au(111) surface},\ }\href {https://doi.org/10.1103/PhysRevE.107.014133} {\bibfield  {journal} {\bibinfo  {journal} {Phys. Rev. E}\ }\textbf {\bibinfo {volume} {107}},\ \bibinfo {pages} {014133} (\bibinfo {year} {2023})}\BibitemShut {NoStop}%
\bibitem [{\citenamefont {Borisenko}\ \emph {et~al.}(2011)\citenamefont {Borisenko}, \citenamefont {Cortese}, \citenamefont {Fiore}, \citenamefont {Gravina},\ and\ \citenamefont {Papa}}]{Borisenko2011}%
  \BibitemOpen
  \bibfield  {author} {\bibinfo {author} {\bibfnamefont {O.}~\bibnamefont {Borisenko}}, \bibinfo {author} {\bibfnamefont {G.}~\bibnamefont {Cortese}}, \bibinfo {author} {\bibfnamefont {R.}~\bibnamefont {Fiore}}, \bibinfo {author} {\bibfnamefont {M.}~\bibnamefont {Gravina}},\ and\ \bibinfo {author} {\bibfnamefont {A.}~\bibnamefont {Papa}},\ }\bibfield  {title} {\bibinfo {title} {Numerical study of the phase transitions in the two-dimensional {Z}(5) vector model},\ }\href {https://doi.org/10.1103/PhysRevE.83.041120} {\bibfield  {journal} {\bibinfo  {journal} {Phys. Rev. E}\ }\textbf {\bibinfo {volume} {83}},\ \bibinfo {pages} {041120} (\bibinfo {year} {2011})}\BibitemShut {NoStop}%
\bibitem [{\citenamefont {Surungan}\ \emph {et~al.}(2019)\citenamefont {Surungan}, \citenamefont {Masuda}, \citenamefont {Komura},\ and\ \citenamefont {Okabe}}]{Surungan2019}%
  \BibitemOpen
  \bibfield  {author} {\bibinfo {author} {\bibfnamefont {T.}~\bibnamefont {Surungan}}, \bibinfo {author} {\bibfnamefont {S.}~\bibnamefont {Masuda}}, \bibinfo {author} {\bibfnamefont {Y.}~\bibnamefont {Komura}},\ and\ \bibinfo {author} {\bibfnamefont {Y.}~\bibnamefont {Okabe}},\ }\bibfield  {title} {\bibinfo {title} {{B}erezinskii–{K}osterlitz–{T}houless transition on regular and villain types of $q$-state clock models},\ }\href {https://doi.org/10.1088/1751-8121/ab226d} {\bibfield  {journal} {\bibinfo  {journal} {J. Phys. A: Math. Theor.}\ }\textbf {\bibinfo {volume} {52}},\ \bibinfo {pages} {275002} (\bibinfo {year} {2019})}\BibitemShut {NoStop}%
\bibitem [{\citenamefont {Okabe}\ and\ \citenamefont {Otsuka}(2025)}]{Okabe2025}%
  \BibitemOpen
  \bibfield  {author} {\bibinfo {author} {\bibfnamefont {Y.}~\bibnamefont {Okabe}}\ and\ \bibinfo {author} {\bibfnamefont {H.}~\bibnamefont {Otsuka}},\ }\bibfield  {title} {\bibinfo {title} {{BKT} transitions of the {XY} and six-state clock models on the various two-dimensional lattices},\ }\href {https://doi.org/10.1088/1751-8121/ada988} {\bibfield  {journal} {\bibinfo  {journal} {J. Phys. A: Math. Theor.}\ }\textbf {\bibinfo {volume} {58}},\ \bibinfo {pages} {065003} (\bibinfo {year} {2025})}\BibitemShut {NoStop}%
\bibitem [{Din()}]{Ding2025}%
  \BibitemOpen
  \href@noop {} {\ }\bibinfo {note} {R. Ding, H. Hu (unpublished).}\BibitemShut {Stop}%
\bibitem [{\citenamefont {Landi}\ \emph {et~al.}(2025)\citenamefont {Landi}, \citenamefont {Russo}, \citenamefont {Sciortino},\ and\ \citenamefont {Valeriani}}]{Landi2025}%
  \BibitemOpen
  \bibfield  {author} {\bibinfo {author} {\bibfnamefont {C.}~\bibnamefont {Landi}}, \bibinfo {author} {\bibfnamefont {J.}~\bibnamefont {Russo}}, \bibinfo {author} {\bibfnamefont {F.}~\bibnamefont {Sciortino}},\ and\ \bibinfo {author} {\bibfnamefont {C.}~\bibnamefont {Valeriani}},\ }\bibfield  {title} {\bibinfo {title} {Self-assembly of active bifunctional {B}rownian particles},\ }\href {https://doi.org/10.1039/D4SM00805G} {\bibfield  {journal} {\bibinfo  {journal} {Soft Matter}\ }\textbf {\bibinfo {volume} {21}},\ \bibinfo {pages} {45} (\bibinfo {year} {2025})}\BibitemShut {NoStop}%
\bibitem [{\citenamefont {Du}\ \emph {et~al.}(2022)\citenamefont {Du}, \citenamefont {Zhang}, \citenamefont {Pearson}, \citenamefont {Ng}, \citenamefont {McEuen}, \citenamefont {Cohen},\ and\ \citenamefont {Brenner}}]{Du2022}%
  \BibitemOpen
  \bibfield  {author} {\bibinfo {author} {\bibfnamefont {C.~X.}\ \bibnamefont {Du}}, \bibinfo {author} {\bibfnamefont {H.~A.}\ \bibnamefont {Zhang}}, \bibinfo {author} {\bibfnamefont {T.~G.}\ \bibnamefont {Pearson}}, \bibinfo {author} {\bibfnamefont {J.}~\bibnamefont {Ng}}, \bibinfo {author} {\bibfnamefont {P.~L.}\ \bibnamefont {McEuen}}, \bibinfo {author} {\bibfnamefont {I.}~\bibnamefont {Cohen}},\ and\ \bibinfo {author} {\bibfnamefont {M.~P.}\ \bibnamefont {Brenner}},\ }\bibfield  {title} {\bibinfo {title} {Programming interactions in magnetic handshake materials},\ }\href {https://doi.org/10.1039/D2SM00604A} {\bibfield  {journal} {\bibinfo  {journal} {Soft Matter}\ }\textbf {\bibinfo {volume} {18}},\ \bibinfo {pages} {6404} (\bibinfo {year} {2022})}\BibitemShut {NoStop}%
\bibitem [{\citenamefont {Wang}\ \emph {et~al.}(2025)\citenamefont {Wang}, \citenamefont {Sun}, \citenamefont {Chen}, \citenamefont {Wang}, \citenamefont {Chen}, \citenamefont {Chen}, \citenamefont {Shuai}, \citenamefont {Yang}, \citenamefont {Jiao},\ and\ \citenamefont {Liu}}]{Wang2025}%
  \BibitemOpen
  \bibfield  {author} {\bibinfo {author} {\bibfnamefont {J.}~\bibnamefont {Wang}}, \bibinfo {author} {\bibfnamefont {Z.}~\bibnamefont {Sun}}, \bibinfo {author} {\bibfnamefont {H.}~\bibnamefont {Chen}}, \bibinfo {author} {\bibfnamefont {G.}~\bibnamefont {Wang}}, \bibinfo {author} {\bibfnamefont {D.}~\bibnamefont {Chen}}, \bibinfo {author} {\bibfnamefont {G.}~\bibnamefont {Chen}}, \bibinfo {author} {\bibfnamefont {J.}~\bibnamefont {Shuai}}, \bibinfo {author} {\bibfnamefont {M.}~\bibnamefont {Yang}}, \bibinfo {author} {\bibfnamefont {Y.}~\bibnamefont {Jiao}},\ and\ \bibinfo {author} {\bibfnamefont {L.}~\bibnamefont {Liu}},\ }\bibfield  {title} {\bibinfo {title} {Hyperuniform networks of active magnetic robotic spinners},\ }\href {https://doi.org/10.1103/z2rp-21xn} {\bibfield  {journal} {\bibinfo  {journal} {Phys. Rev. Lett.}\ }\textbf {\bibinfo {volume} {134}},\ \bibinfo {pages} {248301} (\bibinfo {year} {2025})}\BibitemShut {NoStop}%
\end{thebibliography}%
\clearpage
\appendix
\onecolumngrid
\section*{End Matter}
\setcounter{figure}{0}
\renewcommand{\thefigure}{A\arabic{figure}}
\renewcommand*{\theHfigure}{\thefigure}

{\it Peak positions of the specific heat and more details for densities $\rho_{\rm s}$ in the asymmetric model} ---

\begin{figure}[htbp] 
   \centering
	\includegraphics[width=3.0 in]{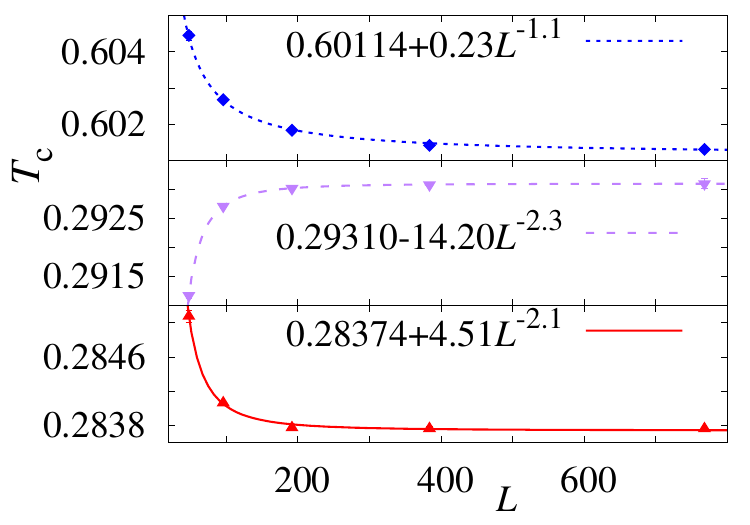}
	\caption{Peak positions $T_c(L)$ of the specific heat versus $L$
	for the asymmetric model. The three subplots from bottom to top correspond to the first to third phase transitions, respectively.} 
   \label{fig:asym-TcL}
\end{figure}

\begin{figure}[htbp] 
   \centering
	\includegraphics[width=3.0 in]{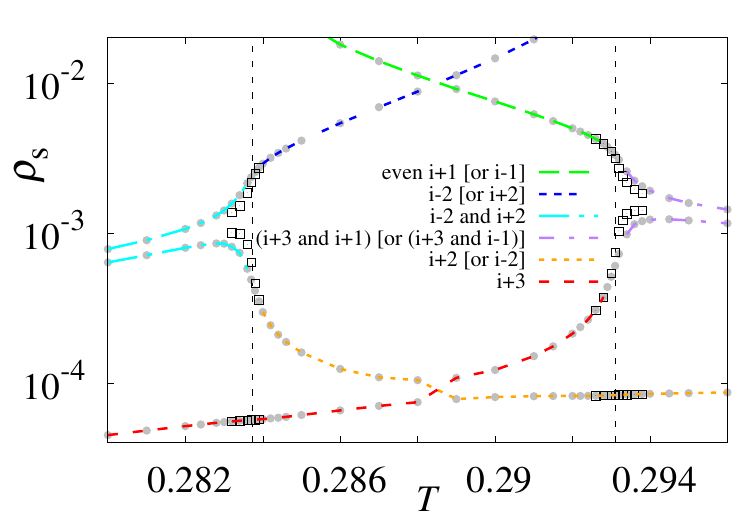}
	\caption{Sorted densities of directors $\rho_{\rm s}$ versus $T$ at $L=384$ (gray dots) 
	and $768$ (black squares) for the asymmetric model. 
   	The plot is an enlargement of the region $\rho_{\rm s}<0.02$ in Fig.3(b) of the main text,
    and all data points below $\rho_{\rm s}=0.01$ are for the fourth to sixth densest director states $\rho_{\rm s}(4:6)$.
    Director states of different curves are given by the labels.
	  Vertical dotted lines indicate positions of the first two phase transitions.
	}
   \label{fig:asym-rhos-enlarge}
\end{figure}
\end{document}